\documentclass{aa}  

\usepackage{graphicx}
\usepackage{txfonts}
\usepackage[switch]{lineno}

\usepackage{todonotes}
\usepackage{graphicx} %
\usepackage{subcaption} %
\usepackage{amsmath} %
\usepackage{booktabs}
\usepackage{multirow}

\usepackage[colorlinks=true, allcolors=blue]{hyperref}
\usepackage[nolist]{acronym}

\usepackage[switch]{lineno}
\usepackage{float}
\usepackage{placeins} %

\begin{document}

   \title{Extracting latent representations from X-ray spectra}
   \subtitle{Classification, regression, and accretion signatures of Chandra sources}

   \author{N. O. Pinciroli Vago
          \inst{1,2,3}
          \and
          R. Martínez-Galarza\inst{3}
          \and
          R. Amato\inst{2}
          }

   \institute{Department of Electronics, Information and Bioengineering, Politecnico di Milano, via G. Ponzio, 34, I-20133 Milan, Italy;\\
              \email{nicolooreste.pinciroli@polimi.it}
         \and
             INAF -- Osservatorio Astronomico di Roma, Via Frascati 33, I-00078, Monte Porzio Catone, Italy;
         \and 
         Center for Astrophysics $\vert$ Harvard \& Smithsonian, 60 Garden Street, Cambridge, MA 02138, USA\\
         \email{jmartine@cfa.harvard.edu}
         }

  \date{Received DD Month YYYY; accepted DD Month YYYY}

  \abstract
  {Spectral signatures are crucial in the era of large X-ray surveys, where effective methods for source identification and classification are needed. Automatic machine learning methods have proven useful in this respect, but so far they have not been applied to large spectral datasets, such as the Chandra Source Catalog (CSC).}
   {This work aims to develop a compact and physically meaningful representation of Chandra X-ray spectra using deep learning. To verify that the learned representation captures relevant information, we evaluate it through classification, regression, and interpretability analyses, and measure the mutual information between spectral and time-domain properties of these sources, aiding in the future identification of transient events.}
   {We use a transformer-based autoencoder to compress X-ray spectra into representations in an 8-dimensional latent space. Astrophysical source types and physical summary statistics are compiled from external catalogs. We evaluate the learned representation in terms of spectral reconstruction accuracy, clustering performance on 8 known astrophysical source classes, and correlation with physical quantities such as hardness ratios and hydrogen column densities ($N_H$).}
   {Upon reconstruction, clustering in the latent space yields a balanced classification accuracy of $\sim$40\% across the 8 source classes, increasing to $\sim$69\% when restricted to AGNs and stellar-mass compact objects exclusively. Moreover, latent features correlate with spectral and temporal properties, suggesting that the compressed representation captures physically relevant information.}
   {Features learned directly from X-ray spectra capture relevant physical information as effectively as human-extracted features that require additional computations. They can be used for both classification and regression in large surveys, and also share mutual information with time-domain properties. The method can be adapted to existing and upcoming X-ray catalogs.}


   \maketitle

   \section{Introduction}
   \label{sec:introduction}

Large catalogs containing hundreds of thousands of astrophysical high-energy sources have been compiled using data obtained by X-ray telescopes, including the Chandra X-ray Observatory \citep{weisskopf_chandra_2000}, XMM-Newton \citep{jansen_xmm-newton_2001}, the Neil Gehrels Swift Observatory \citep{gehrels_swift_2004}, and eROSITA \citep{predehl_erosita_2021}. %
These catalogs contain important diagnostics for the physical characterization of those sources \citep{evans_chandra_2024,webb_xmm-newton_2020,evans_2sxps_2020,merloni_vizier_2024}, such as hardness ratios, fluxes, and variability estimates, next to the reprocessed data products, namely light curves and spectra, that are crucial to identify the nature of the large fraction of serendipitous unlabeled sources.

At the same time, there is a growing interest in systematic searches of X-ray sources with specific spectral properties. For example, in the context of \acp{AGN}, it is generally assumed that the typically hard spectrum of active galaxies results from non-thermal processes in the corona \citep{Galeev+1979ApJ...229..318G}, but a population of soft thermal \acp{AGN} has been identified \citep{Arnaud+1985MNRAS.217..105A,Iwasawa+2024:steep}, a subset of which hosts emerging classes of transient events, such as Tidal Disruption Events \citep[][]{Guolo+2023arXiv230813019G} and Quasi-Periodic Eruptions \citep[][QPEs]{miniutti19,arcodia21,chakraborty21}. Additionally, spectrally hard, fast X-ray transients \citep[][FXTs]{Lin22,quirola22} have been recently associated with neutron star mergers and other accreting compact object phenomena. Many of these sources, especially the extragalactic ones, do not have multi-wavelength counterparts for classification, and therefore it becomes imperative to design effective methods of classification when only X-ray data is available. Complicating the picture, the spectra of astrophysical sources evolve over time, introducing additional variability that is informative of the physical processes but hard to evaluate in the Poisson regime of X-ray photon counting \citep{luo_chandra_2017}. %

Source classification therefore remains a cornerstone of X-ray astronomy, particularly in view of current and forthcoming missions with survey capabilities, such as eROSITA, AXIS \citep{10.1117/12.2310003}, and NewAthena \citep{cruise_newathena_2025}. In particular, understanding to what extent spectral features relate to time-domain anomalies is crucial for the study of AGNs, QPEs, and FXTs. Classification can be attempted with traditional data analysis techniques, typically through spectral modeling aimed at deriving physically meaningful quantities. This approach is employed, for instance, in the \ac{CSC} \citep{evans_chandra_2024}, which collects hundreds of thousands of spectra at different significance levels and provides for each of them a set of relevant summary statistics (e.g., counts, fluxes, best-fit parameters, hardness ratios, variability diagnostics, etc.). In addition to  classification, regression techniques can be used to estimate continuous physical parameters directly from summary statistics. These methods complement traditional spectral modeling by enabling the rapid characterization of source properties across large datasets without the computational cost of individual fitting.

Alternatively, self-supervised approaches can learn representations of astrophysical data across multiple modalities. These techniques have proven effective in diverse fields such as astronomy \citep{orwat-kapola_light-curve_2022, howie_deciphering_2025, rizhko_astrom, cheng_beyond_2021, cheng_identifying_2020}, medicine \citep{huang_self-supervised_2023, yuan_self-supervised_2024}, genomics \citep{richter_delineating_2025}, industrial applications \citep{zangrando_anomaly_2022, xu_application_2025}, and remote sensing \citep{alosaimi_self-supervised_2023, rajaei_self-supervised_2024}. These \ac{ML} methods enable more effective downstream tasks compared to those that require the manual extraction of features, such as classification, regression, and cross-modal prediction of astrophysical quantities (see, e.g., \citealt{parker_astroclip_2024}). Because of their data-driven nature, \ac{ML} algorithms introduce less bias than model-dependent approaches and can expose novel patterns and clusters in the data \citep{duffy_unsupervised_2024}, which, in turn, can provide new insights into physical processes and even help identifying previously unknown astronomical sources.

Autoencoders are a type of \ac{ANN} \citep{yang_autoencoder_2015, kingma_auto-encoding_2022} that learn to compress the input data into a low-dimensional representation (or latent space) in a self-supervised fashion, and reconstruct the input from the compressed latent representation. Autoencoders are particularly useful for anomaly detection \citep{chen_autoencoder-based_2018}, feature extraction \citep{meng_relational_2017}, clustering \citep{tavakoli_autoencoder-based_2020} and data denoising \citep{majumdar_blind_2019}. In astrophysics, they can help identify anomalies in large amounts of data, as done for galaxies \citep{storey-fisher_anomaly_2021,gagliano_physics-informed_2023,liang_autoencoding_2023,arcelin_deblending_2021}, as well as in real-time light curves of different types of objects \citep{muthukrishna_real-time_2022}. Autoencoders can be used to speed up the computation time for parameter estimates 
\citep{tregidga_rapid_2024}, for population studies \citep{saxena_unsupervised_2025}, and to detect outliers based on spectral and variability properties \citep[e.g., the discovery of the extragalactic FXT XRT 200515,][]{dillmann_representation_2025}.

Transformers are another popular architectural choice for sequential data. They process sequences, such as spectra, using a mechanism called attention \citep{vaswani_attention_2017}, that quantifies the level of relatedness between any pair of sequence elements.

In this work, we present a compressed representation of Chandra X-ray spectra, projecting the data in a low-dimensional space using an autoencoder based on a transformer architecture, and use the resulting learned representations to reconstruct the input spectra and classify X-ray sources according to their physical properties.  Specifically, we train our transformer autoencoder to reconstruct the input spectra, cluster the resulting latent space, and evaluate the predictive power of the resulting embeddings on three downstream tasks: X-ray source type classification, regression on physically informative summary statistics and \ac{SR}, to derive mathematical relations between the derived low-dimensional latent space and physical quantities such as X-ray fluxes. 

We evaluate the performance of the learned representations in the classification of 8 types of astronomical objects, and find that, despite the expected degeneracies between astrophysical types, the learned representations have similar performance to manually extracted features at lower computational cost. On the regression front, we demonstrate that the learned features capture physically meaningful spectral summaries such as the hardness ratio and several parameters of physical spectral models, including time variability, stressing their usefulness as compressed summaries of the X-ray spectra, and highlighting mutual information between spectral and time-domain properties. The method can be adapted for the study of X-ray spectra from other facilities, such as XMM-Newton, eROSITA, and other upcoming X-ray missions.

The rest of the paper is organized as follows: Section \ref{sec:data} presents the dataset and data pre-processing; Section \ref{sec:methods} presents the metrics and the algorithms used in this work; Section \ref{sec:results} presents quantitative results; Section \ref{sec:summary} presents the conclusions and discusses potential future directions for research based on this work.

\section{Data}
\label{sec:data}

\subsection{Dataset}
We use a set of background-subtracted X-ray spectra of point-like sources from the \ac{CSC} \citep[v.\,2.1,][]{evans_chandra_2024}. The CSC provides the \ac{PHA} files associated with each source detection, as well as the corresponding  \acp{RMF}\footnote{\url{https://cxc.cfa.harvard.edu/ciao/dictionary/rmf.html}} and \acp{ARF}\footnote{\url{https://cxc.cfa.harvard.edu/ciao/dictionary/arf.html}}, which contain information on the channel-energy conversion and the effective area of the mission, respectively, and are necessary to convert the detected counts into physical units, thus allowing for spectral modeling. We grouped the source spectra using the tools of the \ac{CIAO} software \citep{fruscione_ciao_2006}, with a fixed count binning (i.e., each spectrum is binned differently in energy in order to achieve a fixed value of 10 counts per bin) across all spectra. For context, the spectra of all sources in the \ac{CSC} are fitted using 4 canonical spectral models: power-law, blackbody, bremsstrahlung, and a collisional plasma model \citep[APEC,][]{smith_collisional_2001}. These models were selected by the \ac{CSC} pipeline because they robustly represent the primary X-ray emission mechanisms found in astrophysical sources.

The \ac{CSC} spectra and associated response files used here have been downloaded using the scripts from the Multimodal Universe repository \citep{angeloudi_multimodal_2024}, a large collection of \ac{ML}-ready astronomical datasets, including Chandra. The spectra are filtered based on the number of X-ray counts in the source detection ($N_c \geq 100$), the \ac{SNR} (\acs{SNR} $\geq 5$), and the off-axis angle ($\theta \leq 10$), given that the \ac{PSF} degrades significantly at large angles due to the telescope optics.

In addition to the processed spectra, we also compile a set of labels for each X-ray detection, specifying the astrophysical class of the corresponding source, collected from the labeled dataset prepared by \citet{yang_classifying_2022}: \ac{AGN}, \ac{CV}, \ac{HM-STAR}, \ac{HMXB}, \ac{LM-STAR}, \ac{LMXB}, \ac{NS}, or \ac{YSO}. %

 This results in $\approx 25,000$ spectra, $\approx 3,200$ of which are associated with a known class. We use the subset of labeled X-ray detections as the test set, while the unlabeled data are used for training ($\approx 80\%$ of the unlabeled data) and validation ($\approx 20\%$ of the unlabeled data). Since the class distribution for the training and validation sets is unknown, it is not guaranteed to be the same as the test set. However, the main results presented in this work (e.g., clustering) use only the test set. The test set labels are grouped in two alternative ways, using either the eight distinct classes defined above or only two classes: \acp{AGN} and stellar-mass compact objects, with the latter class comprising \acp{HMXB}, \acp{LMXB}, \acp{CV}, and \acp{NS}. The latter split is done to test whether the automatically learned features can distinguish between \acp{SMBH} and stellar-mass compact objects when only the X-ray spectra are available, as they may display similar spectral properties.

In addition to the astrophysical classes, we collect summary statistics (i.e., the physical quantities reported in Table \ref{tab:tabulated_spectra}) from the \ac{CSC} and fluxes from \citet{yang_classifying_2022}. The summary statistics are relevant for the physical characterization of each detection. We used them to show the correlation with the latent space extracted by the autoencoder (see Sect. \ref{sec:regression}). The fluxes are combined using \ac{SR} (see Sect. \ref{sec:symbolic_regression}). The correlation between fluxes and latent space variables, instead, is not shown, as the spectra are re-scaled and made independent of the absolute flux values.

\subsection{Pre-processing}
\label{sec:preprocessing}

Each spectrum consists of a set of photon count rates (in counts per second per keV) for each energy bin in a range covering from 0.5~keV to 8~keV. Because the spectra have variable bin sizes, whereas the transformer autoencoder needs a fixed-length input, we resample the spectra onto a regular grid of energies via interpolation. Due to the energy-dependent noise level in the proportionality between charge and photon energy in the CCD detector, the energy resolution of Chandra spectra is higher at lower energies. Therefore, we model the spectra using a denser sampling at lower energies and a sparser sampling at higher energies by defining a logarithmic grid in base 10 with $N = 400$ points, spanning from a minimum energy of 0.5\,keV to a maximum energy of 8\,keV. The flux values are linearly interpolated using \texttt{interp1d}\footnote{\url{https://docs.scipy.org/doc/scipy/reference/generated/scipy.interpolate.interp1d.html}}. We determined the optimal number of grid points $N$ by testing values starting at 100 in increments of 50. At each step, the \ac{MAE} between the original and resampled spectra on the test set was computed. The improvement in \ac{MAE} was evaluated using bootstrapping to estimate $5\sigma$ confidence intervals for each step. To balance reconstruction accuracy and the length of the resampled spectra, the procedure was terminated when the $5\sigma$ intervals of consecutive MAE values first overlapped. Linear interpolation was chosen because it acts as a low-pass filter (i.e., it reduces high-frequency components, which also prevents the introduction of additional peaks), as shown in \citet{jokanovic_time-frequency_2015}.

\begin{figure}
    \centering
    \includegraphics[width=\linewidth]{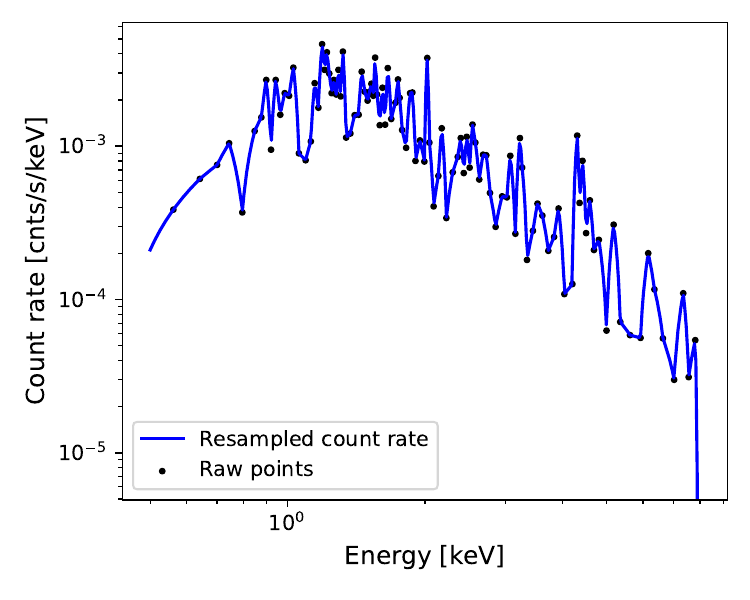}
    \caption{An example of resampling with $N = 400$ points.}
    \label{fig:resampling_example}
\end{figure}

We normalize all the spectra in the count rate dimension to make it fit in the interval $[0,1]$ using min-max normalization, implemented through \texttt{MinMaxScaler}\footnote{\url{https://scikit-learn.org/stable/modules/generated/sklearn.preprocessing.MinMaxScaler.html}}, as shown in Fig. \ref{fig:resampling_example}. For a resampled spectrum $\tilde f$ with flux values $\tilde f_i$, the normalized flux values $\hat f_i$ are given by:

\begin{equation}
    \hat f_i = \frac{\tilde f_i - \min(\tilde f)}{\max(\tilde f) - \min(\tilde f)}
\end{equation}

The goal of normalization is to capture the shape of spectra rather than focusing on their absolute values. Additionally,  normalization of the input has been shown to be beneficial in \ac{ML} \citep{geron_hands-machine_2025}.

\begin{figure}
    \centering
    \includegraphics[width=\linewidth]{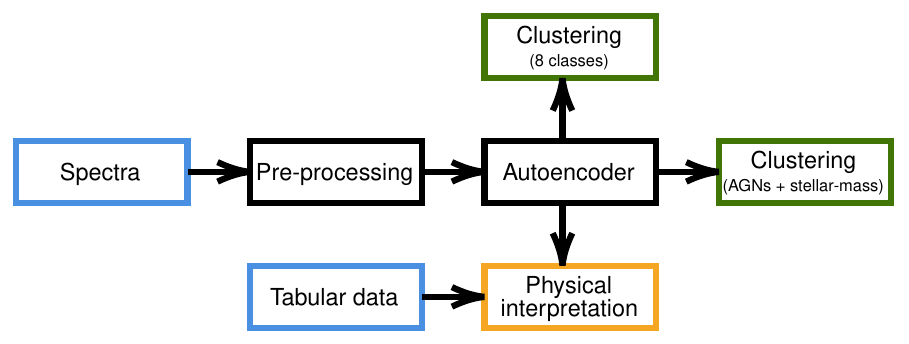}
    \caption{Our pipeline. The blue rectangles indicate the inputs, the green rectangles indicate the clustering outputs, and the orange rectangle indicates the physical interpretation of the results.}
    \label{fig:pipeline}
\end{figure}

\section{Methods}
\label{sec:methods}
This section illustrates the \ac{ML} architectures, our pipeline, and the quantitative evaluation metrics.

\subsection{Pipeline}
Figure \ref{fig:pipeline} presents our pipeline for learning the latent representations, clustering them, using them for source type identification and finally explaining the results through mathematical relations between the latent representations and physical quantities. The pipeline is divided into 6 steps:

\begin{enumerate}
    \item The spectra are pre-processed (Section \ref{sec:preprocessing})
    \item A transformer autoencoder is trained to reconstruct the spectra and to compute a latent representation of the data  (Section \ref{sec:autoencoder}).
    \item The test set latent space vectors are clustered (Section \ref{sec:clustering}) and evaluated in terms of astrophysical class separation.
    \item The clusters are visualized using \ac{tSNE} in two dimensions (Section \ref{sec:tsne})
    \item The physical quantities presented in Table \ref{tab:tabulated_spectra} (excluding fluxes, since we normalised all the spectra) are obtained from the latent space through regression (Section \ref{sec:regression}).
    \item Mathematical relations between fluxes in different energy bands and the latent space dimensions are presented for sources of types \ac{AGN} and \ac{YSO} in the test set (Section \ref{sec:symbolic_regression}).
\end{enumerate}

\subsubsection{Autoencoder}
\label{sec:autoencoder}

\begin{figure}
    \centering
    \includegraphics[width=\linewidth]{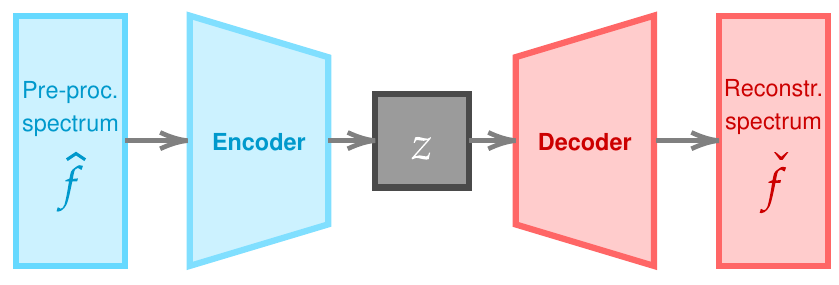}
    \caption{The autoencoder takes in input the resampled normalized spectrum $\hat f$, encodes it into a compressed representation $z$ and reconstructs it as $\check f$.}
    \label{fig:autoencoder}
\end{figure}

We design a deep autoencoder architecture that takes in input the original resampled normalized spectra $\hat f$ and outputs the reconstructed spectra $\check f$, learning a compressed representation (or latent space) of the spectra ($z$), as outlined in Figure \ref{fig:autoencoder}. An overview of the autoencoder architecture is given in Appendix \ref{app:method_autoencoder}.

\begin{table*}
    \centering
    \caption{Overview of the autoencoder hyperparameters, their values, and their meaning.}
    \label{tab:summary_autoencoder}
    \begin{tabular}{l l p{11cm}} 
        \toprule
        \textbf{Hyperparameter} & \textbf{Values} & \textbf{Explanation} \\ 
        \midrule
        \Ac{LR} & $[1\text{e-}4, 5\text{e-}4, 1\text{e-}3]$ & Controls the step size in gradient descent, determining how quickly the model learns. \\ 
        \addlinespace
        Embedding dimensions & $[8, 16]$ & The size of the encoded representation in the latent space. \\ 
        \addlinespace
        Number of layers & $[1, 2, 4]$ & The number of layers in the encoder and decoder sub-networks. \\ 
        \addlinespace
        Number of attention heads & $[8, 16]$ & Number of different attention mechanisms used in each layer. \\ 
        \bottomrule
    \end{tabular}
\end{table*}

Both the encoder and the decoder are implemented using the Transformer architecture \citep{shamsolmoali_vtae_2023, cheng_novel_2024, wu_transformer_2024,vaswani_attention_2017}. Each input sequence of resampled flux values is passed through a linear layer to obtain embedding vectors of dimension $\texttt{embedding\_dim}$. Learnable positional encodings $\mathbf{P}$ of shape $(L, \texttt{embedding\_dim})$ are added to capture information about the ordering of points in the spectra. The resulting sequence is processed by a stack of $\texttt{num\_layers}$ Transformer encoder layers, each employing multi-head self-attention with $\texttt{nhead}$ attention heads (i.e., parallel attention mechanisms that allow the model to focus on different parts of the spectrum at the same time), enabling the model to learn global dependencies across the sequence. After encoding, the sequence is pooled over the length dimension using adaptive average pooling to obtain a single latent vector $\mathbf{z} \in \mathbb{R}^{\texttt{embedding\_dim}}$. For reconstruction, the decoder receives $\mathbf{z}$, applies a linear embedding, adds positional encodings, and passes the result through an analogous stack of Transformer decoder layers. The output layer projects the resulting sequence to produce the reconstructed spectra. The network is optimized using the Adam optimizer \citep{kingma_adam_2017}. The network is trained for a maximum of 200 epochs, with early stopping on the validation \ac{MAE} (patience $p = 20$) to prevent overfitting. Table \ref{tab:summary_autoencoder} provides an overview of the variable hyperparameters, their respective values, and their meanings.

The transformer-based autoencoder is chosen over other autoencoder architectures because transformers are  effective at modeling long-range correlations. In X-ray spectra, such correlations extend beyond local dependencies that are typical of time series data. The self-attention mechanism within transformers allows the model to efficiently identify and represent these non-local relationships across the entire input. In astrophysical terms, this allows the autoencoder to capture physical relationships between spectral features separated by large energy gaps.

\subsubsection{Clustering}
\label{sec:clustering}

Clustering is an unsupervised learning technique used to group similar data points, so that the point in each cluster are more similar to each other than to points in other clusters \citep{jain_data_1999, saxena_review_2017, fotopoulou_review_2024}. In this work, we use clustering to find associations between X-ray spectra in the learned latent space representations for the validation set and evaluate the usefulness of the representation for the two classification tasks introduced in Sect. \ref{sec:data}. We compare 8 clustering algorithms belonging to two families: hard clustering and soft clustering (see also \autoref{tab:summary_cluster}, where each algorithm is labelled according to its clustering type). In hard clustering, each data point is assigned to exactly one cluster, meaning there is no overlap in membership between clusters \citep{jain_data_1999}. In contrast, soft clustering (or fuzzy clustering) assigns each data point to multiple clusters, with membership probabilities that reflect the degree of association with each cluster \citep{jain_data_1999}.  Appendix \ref{app:clustering_methods} presents the clustering algorithms in more detail.

\subsubsection{\acf{tSNE}}
\label{sec:tsne}

The goal of \ac{tSNE}\footnote{\url{https://scikit-learn.org/stable/modules/generated/sklearn.manifold.TSNE.html}} \citep{van_der_maaten_visualizing_2008} is to find a low-dimensional representation (in this work, bi-dimensional) of high-dimensional data (here, the latent space created by the autoencoder) that preserves local pairwise similarities between the original points. Unlike \ac{PCA} \citep{greenacre_principal_2022}, \ac{tSNE} can model complex and non-linear relationships, making it suitable for exploring structures in high-dimensional data. Appendix \ref{app:tsne} presents the algorithm in detail.

\subsubsection{Regression}
\label{sec:regression}

Regression is a technique used to quantify the relationship between input variables and one or more continuous target variables. %
In this work, we use Huber regression \citep{owen_robust_2007} due to its robustness to outliers. This regression technique uses a piecewise loss function that is quadratic for small residuals and linear for large ones, being robust to outliers as a result. To ensure robustness and generalization, we use 5-fold cross-validation. First, the labelled test set is partitioned into five subsets. Then, for each fold, the model is trained on four subsets and validated on the remaining one, with performance metrics (here, \ac{MAE}, see Sect. \ref{sec:metrics}) averaged across all folds. The results are computed first on the entire test set and then by grouping observations according to their associated statistics (see Table~\ref{tab:tabulated_spectra}). In this second case, for each observation, the best-fitting model (either black body, power law, bremsstrahlung, or collisional plasma) is selected, and only the quantities associated with it are estimated.

\subsubsection{\Acf{SR}}
\label{sec:symbolic_regression}

\ac{SR} \citep{makke_interpretable_2024} searches for underlying mathematical expressions that describe the relationships between variables. Unlike traditional regression methods, which require assumptions about the form of the underlying model \citep{maulud_review_2020}, \ac{SR} is able to discover adequate expressions that combine variables autonomously. The expressions are represented as trees. The nodes in the trees represent either operators and functions (e.g., +, -, $\times$, $\div$, $\sin$, $\exp$) or variables and constants. In this work, \ac{SR} is used to find relations between the latent space components and the fluxes presented in Table \ref{tab:tabulated_spectra} using 4 operations (+, -, $\times$, $\div$) and a function ($\log$). The use of $\log$ is motivated by the fact that flux values span several orders of magnitude, making log-transformation a natural choice. From the extracted formulas, we manually derive simplified versions proportional to the original ones to identify simple correlations with the latent space dimensions. Appendix \ref{app:symbolic_regression} presents the \ac{SR} algorithm in more detail.

\subsection{Metrics}
\label{sec:metrics}

The results are evaluated in three steps, explained in more detail in the following subsections:
\begin{itemize}
    \item Autoencoder reconstruction: assessing the reconstruction ability of the autoencoder (see Sect. \ref{sec:metrics_ae}).
    \item Clustering performance: evaluating the clustering results on the test set for both the 8-class and 2-class cases (see Sect. \ref{sec:metrics_clustering}).
    \item Physical interpretation of the latent space: for all the sources, evaluating the relationship between the latent space and physical quantities; for \acp{AGN}, evaluating the correlation between physical interpretable formulas and the latent space dimensions (see Sect. \ref{sec:metrics_sr}).
\end{itemize}

\subsubsection{Autoencoder}
\label{sec:metrics_ae}
The reconstruction ability of the autoencoder is quantified using the \ac{MAE} between the resampled normalized spectra and their corresponding reconstructions. For a given resampled and normalized spectrum $\hat f$ with count rates $\hat f_i$ and estimated count rates $\check f_i$, the \ac{MAE} is defined as:

\begin{equation}
    \ac{MAE} = \frac{1}{n}\sum_{i=1}^n |\check f_i - \hat f_i|
\end{equation}
where $n$ represents the number of points in the spectrum. %
The \ac{MAE} ranges between 0 and 1, with lower values indicating a better reconstruction ability (i.e., a better autoencoder performance).
Unlike \ac{MSE}, which squares errors and effectively amplifies outliers, \ac{MAE} applies a linear penalty. This makes \ac{MAE} significantly more robust, as extreme values do not dominate the metrics. It is also easier to interpret, as it measures average relative error on the normalized spectra. Moreover, it avoids the bias \ac{MSE} would have in high-flux regions, since squaring errors in \ac{MSE} emphasizes deviations at large count values. On the other hand, \ac{MAE} is less smooth than \ac{MSE}, which can lead to slower or less stable optimization.

\subsubsection{Clustering}
\label{sec:metrics_clustering}

The clustering performance is evaluated on the labelled test set using a contingency matrix \citep{bonaccorso_hands-unsupervised_2019}. A contingency matrix is a tool used to compare the correspondence between \ac{GT} labels and the predicted labels (i.e., clusters) produced by the clustering algorithm. In a generic clustering problem, let $m$ denote the number of \ac{GT} classes $\{C_1, C_2, ..., C_m\}$ and $k$ denote the number of predicted clusters $\{K_1, K_2, ..., K_k\}$. For this study, the number of clusters is fixed such that $k = m = 8$ (for the algorithms that allow to do so, see Appendix \ref{app:clustering_methods}). 

In the contingency matrix $M$, each row corresponds to a \ac{GT} class, and each column corresponds to a cluster. A cell $M_{ij}$ indicates the number (or proportion) of samples from class $C_i$ assigned to cluster $K_j$. In this work, the contingency matrix is adjusted for each class, yielding the normalized matrix $M'$, defined as:

\begin{equation}
    M'_{ij} = \frac{M_{ij}}{\sum_{j=1}^k M_{ij}} \quad \forall i, j
\end{equation}

As a result, the sum of the elements in each row of the normalized matrix is equal to 1:

\begin{equation}
    \sum_{j=1}^k M'_{ij} = 1 \quad \forall i
\end{equation}

The evaluation procedure is formed by two steps consists of the computation of the contingency matrix, followed by the assignment of each cluster to a corresponding \ac{GT} class. The assignment of clusters to \ac{GT} classes is determined to maximize the trace of the normalized contingency matrix, $Tr(M')$, which is the sum of the elements on its main diagonal. Maximizing $Tr(M')$ directly corresponds to maximizing the balanced accuracy \citep{grandini_metrics_2020}, defined as:

\begin{equation}
    ACC_B = \frac{1}{k}\sum_{i=1}^{k} M'_{ii} = \frac{Tr(M')}{k}
\end{equation}

The optimal assignment of clusters to classes is computed using the Kuhn-Munkres algorithm, as implemented in \texttt{scipy.optimize.linear\_sum\_assignment}\footnote{\url{https://docs.scipy.org/doc/scipy/reference/generated/scipy.optimize.linear\_sum\_assignment.html}} \citep{crouse_implementing_2016}.

For the best balanced accuracy, the uncertainty is computed using bootstrapping on the clusters, with a 99\% confidence interval and $n = 10000$ bootstrapping iterations. Uncertainties obtained using bootstrapping account for class imbalance and the limited number of test samples, allowing a fair comparison of the results. For simplicity and compactness, the other results are reported without uncertainties.

\subsubsection{Regression and \acl{SR}}
\label{sec:metrics_sr}

The regression results are evaluated using the \ac{MAE}, consistent with the autoencoder case. Performance is compared against a baseline, defined as the mean value of the target variable, and the improvement is reported as a percentage.

The evaluation of \ac{SR} formulas is performed using two correlation coefficients: the Pearson correlation coefficient $\rho$\footnote{\url{https://docs.scipy.org/doc/scipy/reference/generated/scipy.stats.pearsonr.html}}, which measures linear correlations,  and the Spearman correlation coefficient $r_s$\footnote{\url{https://docs.scipy.org/doc/scipy/reference/generated/scipy.stats.spearmanr.html}}, which assesses rank-based correlations \citep{martinez-gil_novel_2020}.

The Pearson correlation coefficient is defined as:

\begin{equation}
    \rho = \frac{1}{n-1} \sum_{i=1}^n \frac{(X_i - \bar X)(Y_i - \bar Y)}{s_X s_Y}
\end{equation}

The Spearman correlation coefficient is defined as:

\begin{equation}
    r_s = \frac{1}{n}\sum_{i=1}^n \frac{X_i Y_i - \bar X \bar Y}{s_X s_Y}
\end{equation}

In both cases, $n$ represents the number of samples, $X_i$ and $Y_i$ represent the values associated with the $i$-th sample (i.e. the latent space variable and the physical formula value),  $\bar{X}$ and $\bar{Y}$ are their respective means, and $s_X$ and $s_Y$ are their standard deviations.    Both $\rho$ and $r_s$ range from -1 to 1, with larger absolute values indicating stronger correlations.

\section{Results and discussion}
\label{sec:results}

We now report the results of applying the transformer autoencoder framework to our dataset of Chandra spectra. We first evaluate the reconstruction capabilities of the autoencoder. We then report on the classification and regression tasks, and investigate interpretability through symbolic regression.

\subsection{Spectral reconstruction}

\begin{figure*}
    \centering
    \includegraphics[width=\linewidth]{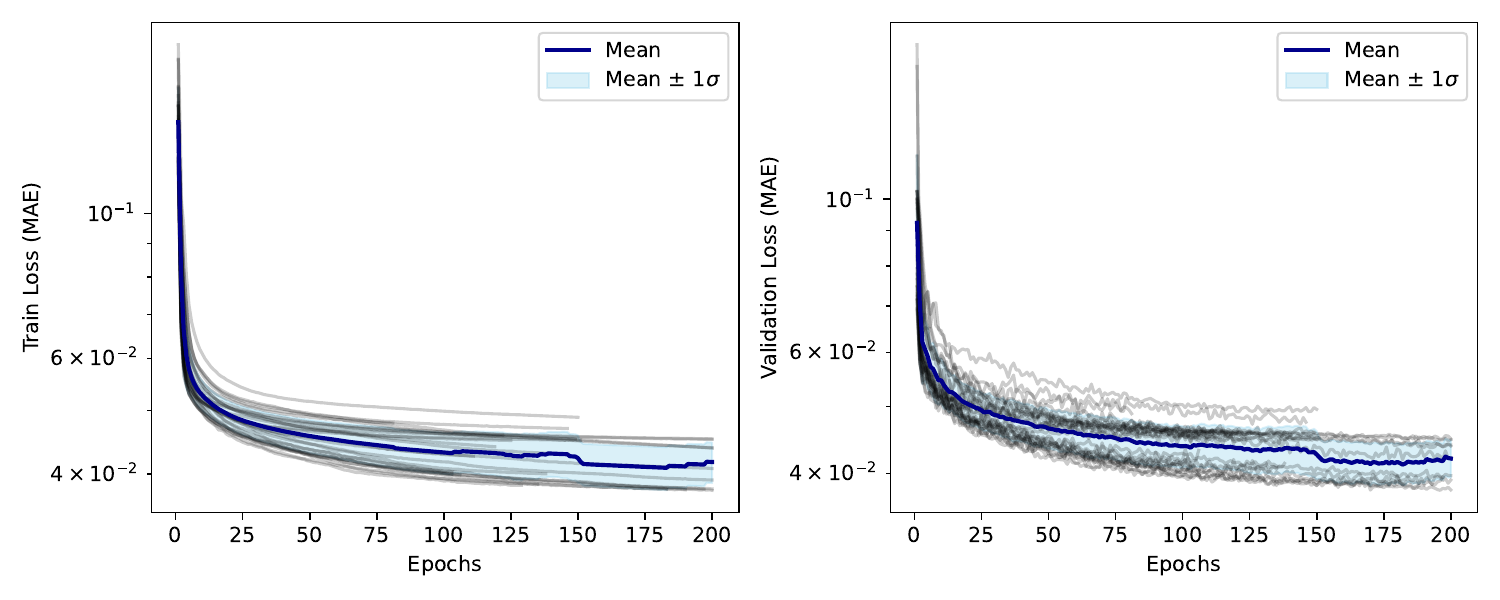}
    \caption{The training (left panel) and validation (right panel) losses of the autoencoder, measured using \ac{MAE}. The dark blue line is the mean loss across different hyperparameter configurations in the grid search (see \autoref{sec:autoencoder}), the light blue region indicates the mean $\pm 1\sigma$ and the gray lines indicate the loss for different hyperparameters configurations. The number of epochs varies across the experiments because of early stopping.}
    \label{fig:autoencoder_losses}
\end{figure*}

\begin{figure*}
    \centering
    \begin{subfigure}[t]{0.33\linewidth}
        \centering
        \includegraphics[width=\linewidth]{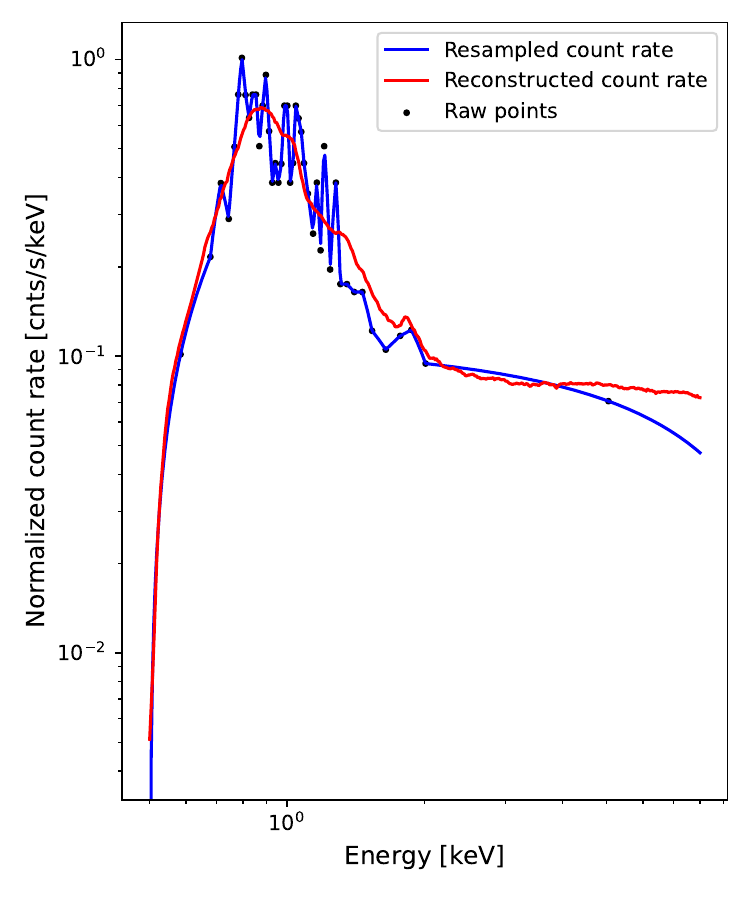}
        \label{fig:subfig1}
    \end{subfigure}
    \hfill
    \begin{subfigure}[t]{0.33\linewidth}
        \centering
        \includegraphics[width=\linewidth]{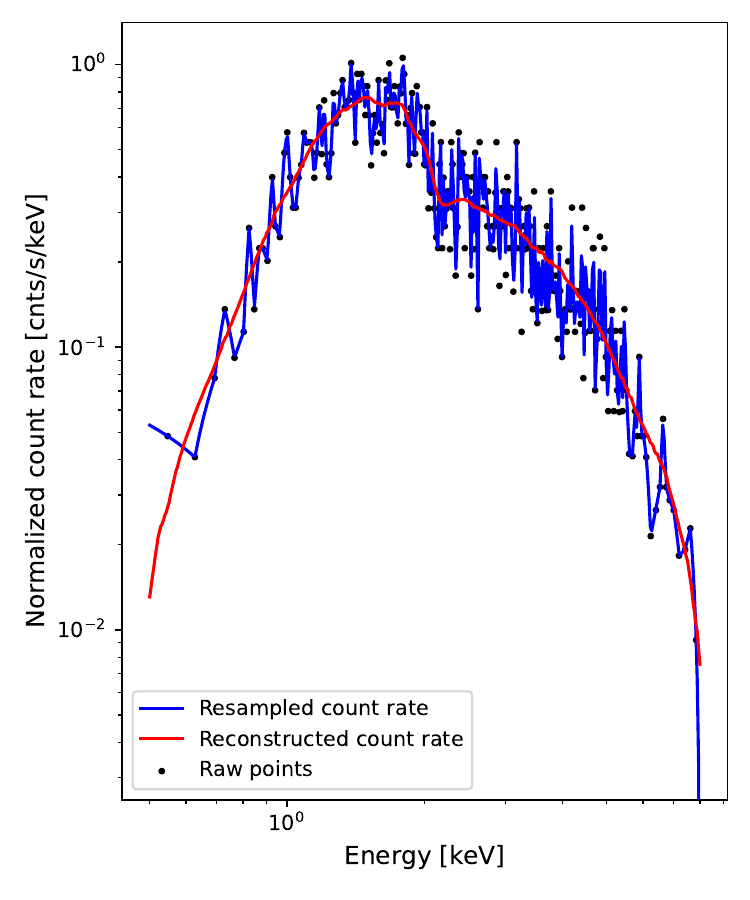}
        \label{fig:subfig2}
    \end{subfigure}
    \hfill
    \begin{subfigure}[t]{0.33\linewidth}
        \centering
        \includegraphics[width=\linewidth]{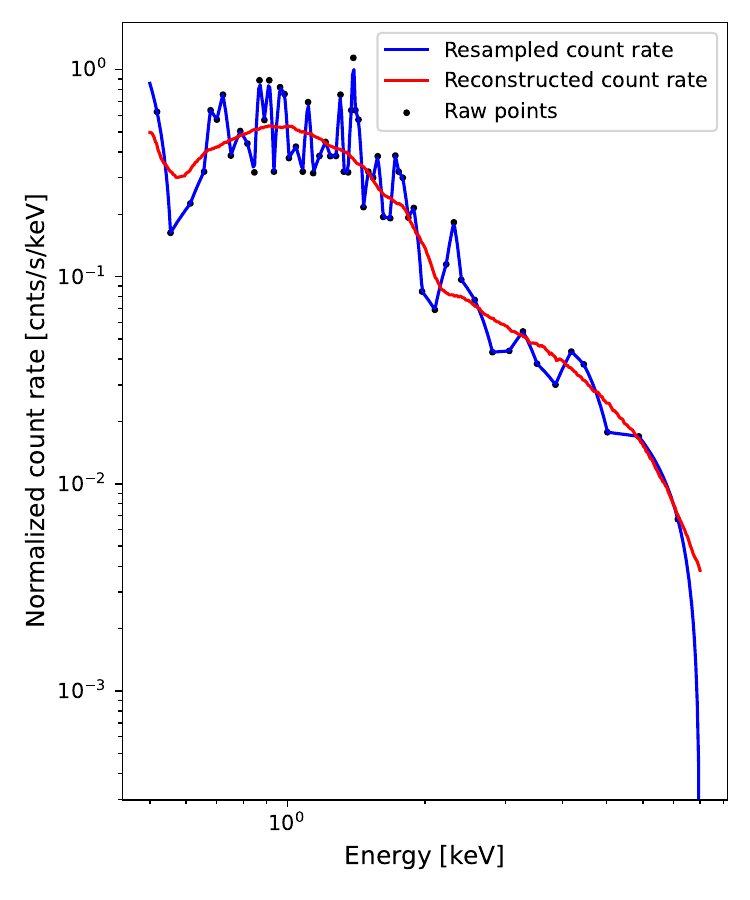}
        \label{fig:subfig3}
    \end{subfigure}
    \caption{Examples of autoencoder reconstructions (red) compared to the input resampled normalized spectra (blue). 
    Left: the autoencoder reconstructs the fluxes for lower energy values better (source: 2CXO J034639.3+240611, Obs. ID: 17250, type: \ac{LM-STAR}). 
    Centre: the autoencoder reconstructs the fluxes for higher energy values better (source: 2CXO J111438.8+324133, Obs. ID: 3137, type: \ac{AGN}). 
    Right: the autoencoder reconstruction capabilities are comparable for the fluxes at all energies (source: 2CXO J100433.8+411234, Obs. ID: 14498, type: \ac{AGN}).}
    \label{fig:smoothed}
\end{figure*}

Figure \ref{fig:autoencoder_losses} shows the training and validation losses for the autoencoder. A non-zero loss suggests that the autoencoder reconstructs a smoothed version of the input spectra, as illustrated in Fig. \ref{fig:smoothed}. The use of a logarithmic scale on the y-axis and the different scales in the three spectra count ranges visually accentuate differences between the reconstructed and the original spectra, which appear larger at lower count rates. The reconstruction in the left panel has $\ac{MAE} \approx 0.0337$, the one in the central panel has $\ac{MAE} \approx 0.0523$ and the one in the right panel has $\ac{MAE} \approx 0.0612$. In the presented examples, the largest errors correspond to regions with more fluctuations in the count rate values.
In all hyperparameter configurations, the losses on both the train and validation sets decrease, indicating the absence of overfitting. All the configurations result in a validation set \ac{MAE} < 0.06. Overall, the autoencoder's performance is consistent across different hyperparameter settings, with a relatively small spread in performance, suggesting that the model is robust to hyperparameter choices. In all experiments, both the training and validation losses decrease rapidly within the first $\approx 25$ training epochs.

\begin{figure*}[htbp]
\centering
\begin{subfigure}{0.32\textwidth}
    \includegraphics[width=\linewidth]{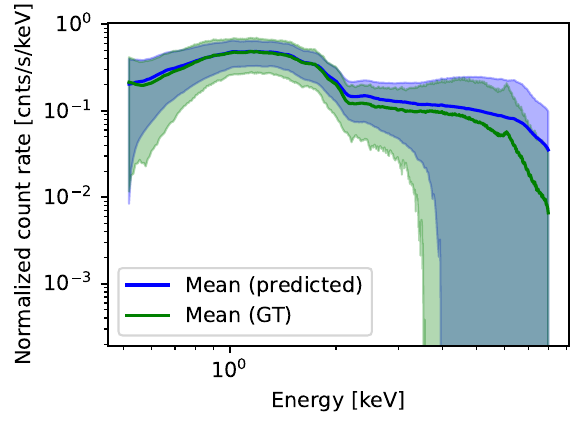}
    \caption{\ac{AGN}}
    \label{fig:sAGN}
\end{subfigure}
\begin{subfigure}{0.32\textwidth}
    \includegraphics[width=\linewidth]{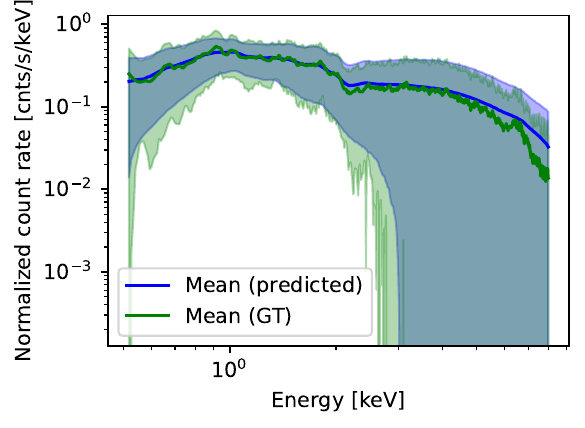}
    \caption{\ac{CV}}
    \label{fig:sCV}
\end{subfigure}
\begin{subfigure}{0.32\textwidth}
    \includegraphics[width=\linewidth]{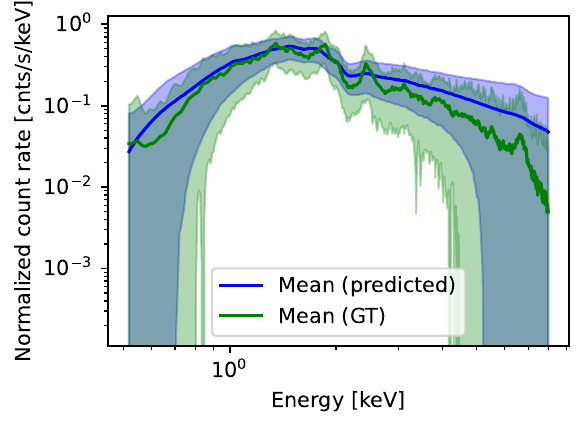}
    \caption{\ac{HM-STAR}}
    \label{fig:sHM_STAR}
\end{subfigure}
\begin{subfigure}{0.32\textwidth}
    \includegraphics[width=\linewidth]{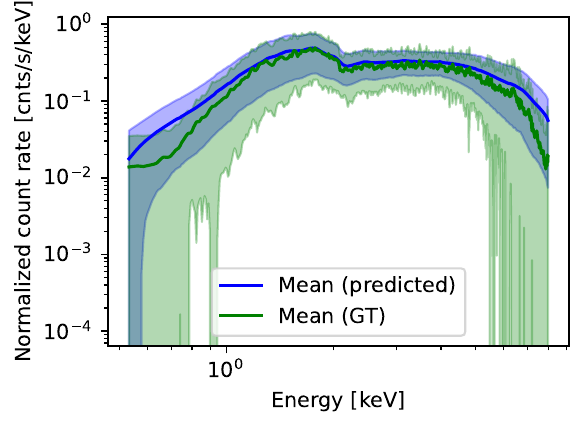}
    \caption{\ac{HMXB}}
    \label{fig:sHMXB}
\end{subfigure}
\begin{subfigure}{0.32\textwidth}
    \includegraphics[width=\linewidth]{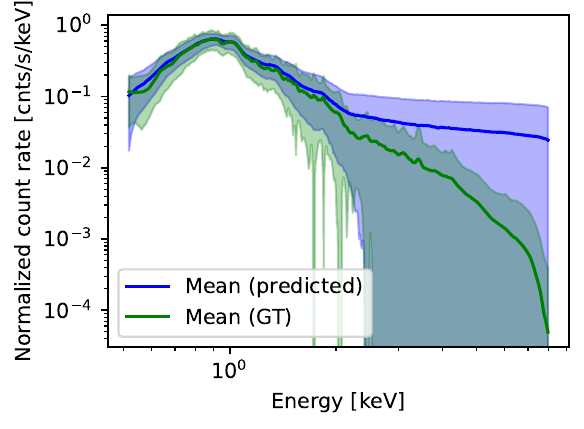}
    \caption{\ac{LM-STAR}}
    \label{fig:sLM_STAR}
\end{subfigure}
\begin{subfigure}{0.32\textwidth}
    \includegraphics[width=\linewidth]{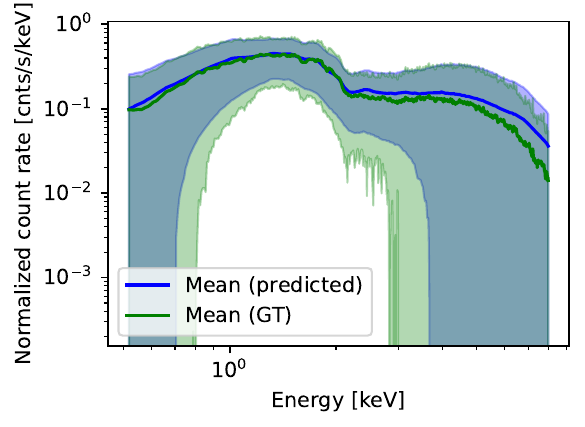}
    \caption{\ac{LMXB}}
    \label{fig:sLMXB}
\end{subfigure}
\begin{subfigure}{0.32\textwidth}
    \includegraphics[width=\linewidth]{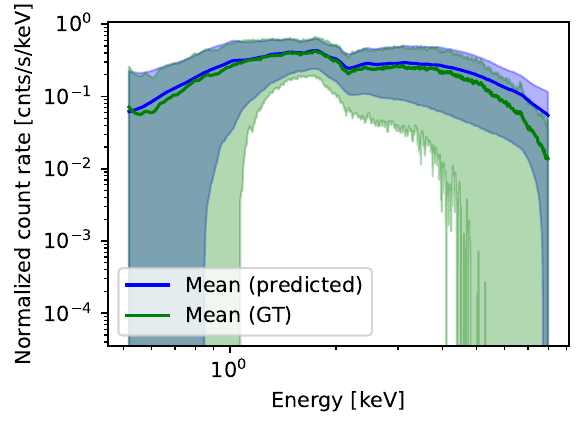}
    \caption{\ac{NS}}
    \label{fig:sNS}
\end{subfigure}
\begin{subfigure}{0.32\textwidth}
    \includegraphics[width=\linewidth]{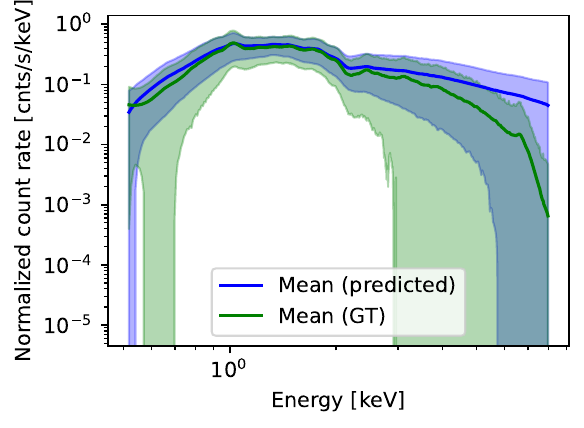}
    \caption{\ac{YSO}}
    \label{fig:sYSO}
\end{subfigure}

\caption{Comparison of the mean spectra (green) and the corresponding reconstructed spectra (blue) for each class. The shaded regions represent the 1$\sigma$ standard deviation around the mean.}
\label{fig:smoothed_collective}
\end{figure*}

Fig. \ref{fig:smoothed_collective} shows the comparison, on the labelled test set and for each class, between the original spectra and the reconstructed spectra. The solid green line is the mean value for the original spectra, and the solid blue line is the mean value for the spectra reconstructed by the autoencoder. The shaded regions indicate a $\pm 1\sigma$ standard deviation around the mean. Overall, the reconstruction means are better for lower energies ($E \lesssim 3$ keV), and the original spectra mean values are within the $1\sigma$ intervals around the reconstructed spectra mean values. The most significant differences are observed for \acp{LM-STAR} and \acp{YSO}. Both classes are characterized by lower normalized count rates at higher energies. In addition, \acp{LM-STAR} are rare (96 out of 3241, or $\approx 3\%$). Still, the average reconstruction at high energies lies  within 1$\sigma$ of the average ground truth spectra. Reconstruction quality does not depend directly on the class since the autoencoder is unsupervised. However, since the autoencoder is predominantly trained on hard spectra (emission dominates in the 2--8\,keV range), it implicitly learns to expect higher fluxes at high energies, which in turn leads to poor generalisation on this class, whose soft spectra deviate from the norm. 

\subsection{Classification}

\begin{figure*}
    \centering
    \includegraphics[width=0.85\textwidth]{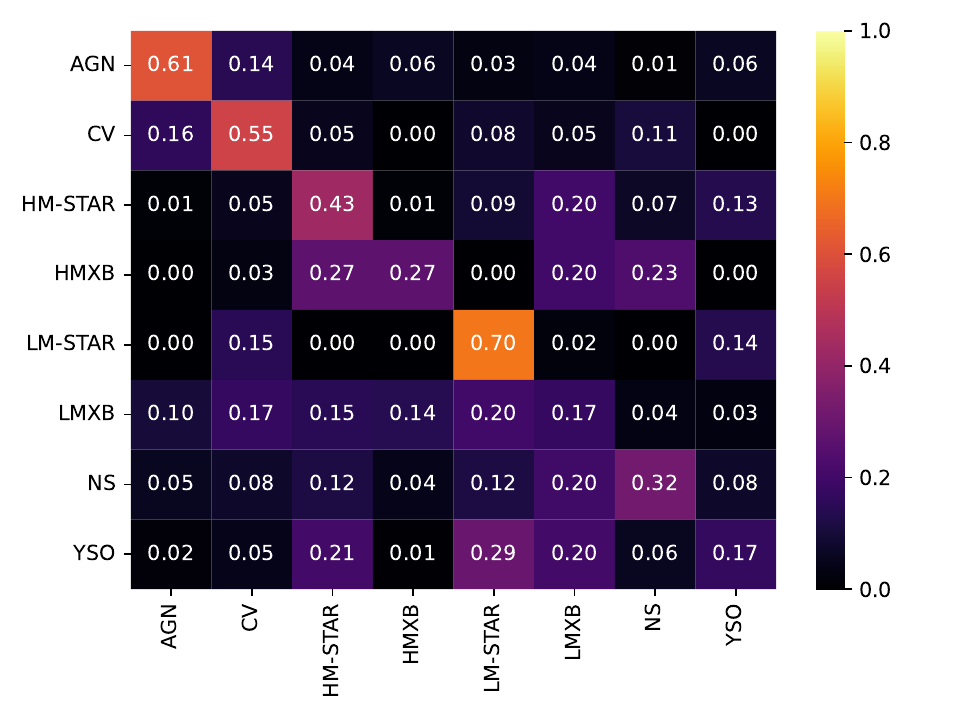}
    \caption{The best contingency matrix, showing a balanced accuracy of $\approx 40\%$. The \ac{GT} classes are shown on the y-axis, and the predicted classes are shown on the x-axis.}
    \label{fig:heatmap}
\end{figure*}

Figure \ref{fig:heatmap} shows the contingency matrix for the best hyperparameters' configuration (\ac{LR} of $10^{-4}$, 8 embedding dimensions, 2 layers, 8 attention heads, and a \ac{GMM} with a 'full' covariance type). The balanced accuracy is $\approx 40.17\%$, significantly higher than the $12.5\%$ expected from an uninformed random classifier. \ac{GMM} outperforms alternative algorithms because, unlike distance-based methods, it assigns probabilistic memberships that reflect that there is no one-to-one correspondence between spectral characteristics and source classes. The 'full' covariance type allows each mixture component to capture distributions better aligned with continuous physical properties (e.g., hardness ratios), which the autoencoder captures along latent dimensions. Appendix \ref{app:results_clustering} presents a thorough comparison of different clustering algorithms and autoencoder hyperparameters.

\subsubsection{Spectral degeneracies}
\label{sec:degeneracies}
Our pipeline performs particularly well in classifying \acp{AGN} and \acp{LM-STAR}. Interestingly, although \ac{LM-STAR} spectra are not well reconstructed, their reconstruction patterns are still clearly distinguishable from those of other spectra. The lowest performance, instead, is observed for \acp{HMXB}, \acp{LMXB}, and \acp{YSO}. Note that the physical grounding of the latent space is further supported by the symbolic regression analysis (\autoref{sec:physical_correlations}) and the regression results (\autoref{sec:results_regression}), which we refer to throughout this discussion.

To assess the significance of differences between classes we compute, for each resampled energy bin in the spectra:
\begin{equation}
    m = \frac{|\mu_{A}-\mu_{B}|}{\sqrt{\sigma_{A}^2+\sigma_{B}^2}}
\end{equation}
where $A$ and $B$ represent the two classes under comparison. Higher values of $m$ indicate significant differences between the spectral shapes of the two classes, while lower values suggest that the spectra are  similar. If $m < 1$ for the majority of energy bins, the spectral differences between the classes are not considered statistically significant.

Based on the contingency matrix, we identify the following main sources of confusion:

\begin{itemize}
    \item \acp{YSO} and \acp{LM-STAR}: in \acp{YSO}, X-ray emission arises from a combination of processes: particles accelerated by strong rotation-powered magnetic fields, heating of the corona, heating of the disk due to accretion, and shocks in outflows and jets. In contrast, \acp{LM-STAR} (typically old, main-sequence stars) exhibit X-ray emission primarily due to coronal heating, also linked to magnetic activity from rotation, but are not determined by accretion or jets.  We notice statistically significant differences between spectra ($> 1\sigma$) only for $E \lesssim 2$\,keV. Overall, $m < 1$ for $\approx 78\%$ of the points in the predictions and $\approx 86\%$ in the \ac{GT}. 

    \medskip

    \item \acp{YSO} and \acp{HM-STAR}: The spectra of the two classes also have $m < 1$ for 100\% of the energy bins. Massive \acp{YSO} drive powerful stellar winds whose X-ray signatures overlap with those of \acp{HM-STAR}, which also produce hard emission from stellar-wind shocks and magnetic activity, making the two classes spectrally similar at the coarse energy resolution considered here.
    \item \acp{HMXB} and \acp{HM-STAR}: an \ac{HM-STAR} is the massive stellar companion in an \ac{HMXB} system \citep{tan_high-mass_2021, ge_adiabatic_2024}. In \acp{HMXB}, the X-ray emission is due to accretion, including wind interactions and magnetic activity. In isolated \acp{HM-STAR}, X-rays are generated mainly by stellar winds and magnetic processes. Since the autoencoder latent variables are correlated with physical variables rather than directly with classes (cf Section \ref{sec:regression}), the clustering algorithm is not able to fully separate the spectra associated with different classes. Moreover, the two classes have $m < 1$ for 84\% of the energy bins in predicted data and for 90\% of the energy bins in \ac{GT} data. The only significant difference between spectra is observed for $E \approx 4$\,keV.
    \item \acp{AGN} and \acp{CV}: approximately 14\% of \acp{AGN} are classified as \acp{CV}, and about 16\% of \acp{CV} are classified as \acp{AGN}, suggesting spectral similarities notwithstanding the different physical nature of the two classes. In both cases, accretion is the primary mechanism of X-ray emission, producing both thermal radiation from hot disks and coronae and non-thermal emission from accelerated particles. The spectra of the two classes also have $m < 1$ for 100\% of the energy bins. %
    \item \acp{LM-STAR} and \acp{CV}: approximately 15\% of \acp{LM-STAR} are mispredicted as \acp{CV}. 
    This is physically motivated by the fact that the X-ray emission in CVs and low-mass stars is a thermal plasma, with similar temperatures, though the nature of the systems is fundamentally different. In CV's the thermal plasma comes from the accretion flow between the companion star and the white dwarf, whereas in low-mass stars thermal emission comes from the corona. The spectra of the two classes also have $m < 1$ for 100\% of the points in the predictions and 98\% in the \ac{GT}, which implies that X-ray spectral signatures are insufficient to discern between the two.
    \item \acp{HM-STAR} and \acp{LMXB}: both are characterized by relatively hard spectra compared to other classes (e.g., \acp{LM-STAR}). Moreover, the spectra of the two classes also have $m < 1$ for 100\% of the energy bins.
    \item \acp{YSO} and \acp{LMXB}: approximately 20\% of \acp{YSO} are mispredicted as \acp{LMXB}. Both classes can produce hard X-ray emission, resulting in overlapping spectral shapes. The two classes have $m < 1$ for 100\% of the energy bins, indicating similar spectra
    
    \item \acp{HMXB} and \acp{NS}: some confusion is observed between these two classes. In general, HMXB are expected to have a distinct hard X-ray emission related to Comptonization associated with accretion, whereas neutron stars, if truly isolated, are characterized by thermal blackbody emission. However, if the HMXB is in a low-accretion, quiescent state, thermal emission from the disk can dominate. Similarly, HMXB can also be heavily absorbed in embedded in strong winds, which can distort its rich spectrum and make it appear featureless. Both of these effects can be exacerbated at low signal to noise.
    
\end{itemize}

\subsubsection{Clustering}

\begin{figure*}[htbp]
\centering
\begin{subfigure}{0.32\textwidth}
    \includegraphics[width=\linewidth]{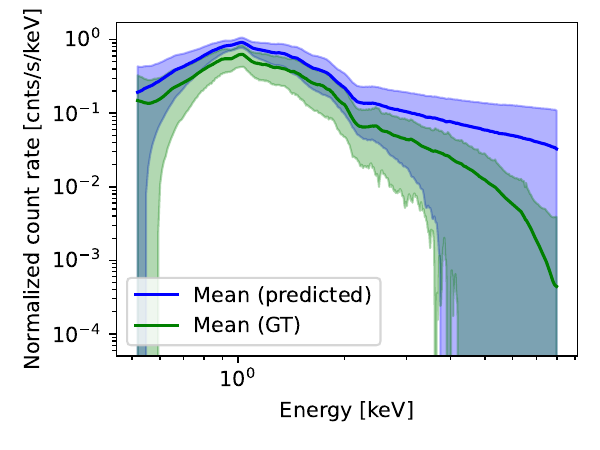}
    \caption{-1 $\leq$ \texttt{hard\_hs} < -0.33}
\end{subfigure}
\begin{subfigure}{0.32\textwidth}
    \includegraphics[width=\linewidth]{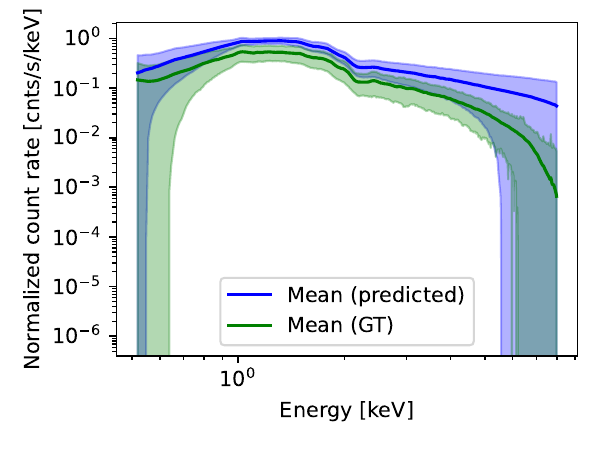}
    \caption{-0.33 $\leq$ \texttt{hard\_hs} < 0.33}
\end{subfigure}
\begin{subfigure}{0.32\textwidth}
    \includegraphics[width=\linewidth]{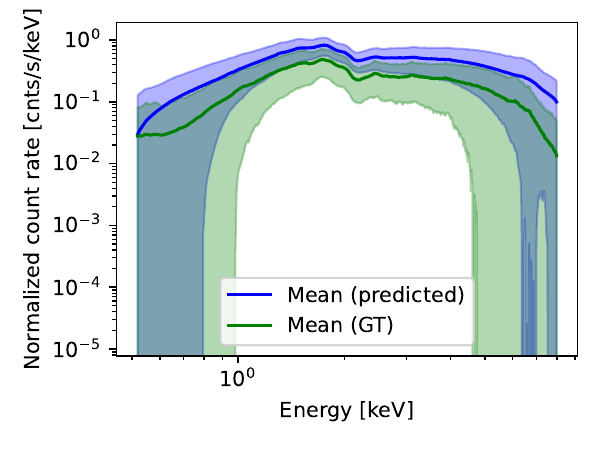}
    \caption{0.33 $\leq$ \texttt{hard\_hs} < 1} 
\end{subfigure}
\caption{Comparison of mean reconstructed (blue) and ground-truth (green) spectra for three \texttt{hard\_hs} ranges: (a) soft spectra ($-1 \leq \texttt{hard\_hs} < -0.33$), (b) intermediate spectra ($-0.33 \leq \texttt{hard\_hs} < 0.33$), and (c) hard spectra ($0.33 \leq \texttt{hard\_hs} \leq 1$). Solid lines show the mean spectrum across all sources in each bin, and shaded regions indicate the $1\sigma$ standard deviation around the mean.}
\label{fig:hardness_slice}
\end{figure*}

\begin{figure*}[htbp]
    \centering
    \mbox{
    \begin{subfigure}{0.32\textwidth}
        \includegraphics[width=\textwidth]{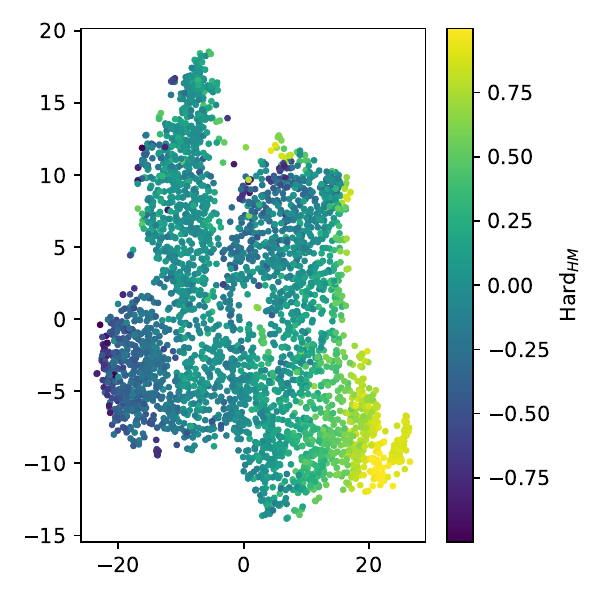}
        \caption{\texttt{hard\_hm}}
        \label{fig:hard_hm}
    \end{subfigure}%
    \begin{subfigure}{0.32\textwidth}
        \includegraphics[width=\textwidth]{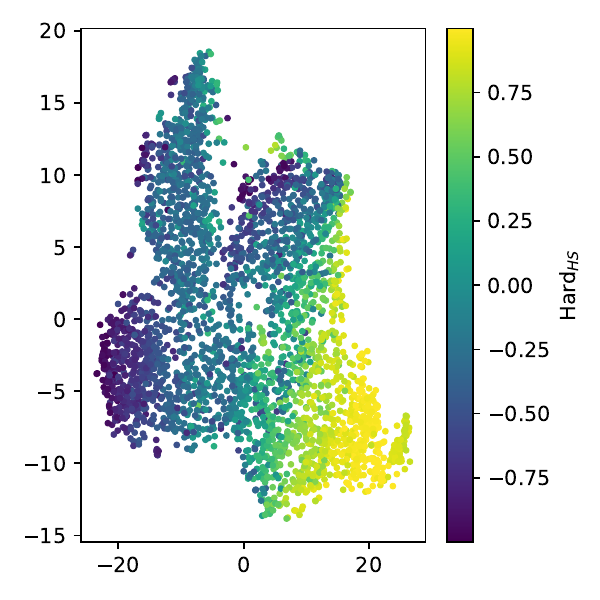}
        \caption{\texttt{hard\_hs}}
        \label{fig:hard_hs}
    \end{subfigure}%
    \begin{subfigure}{0.32\textwidth}
        \includegraphics[width=\textwidth]{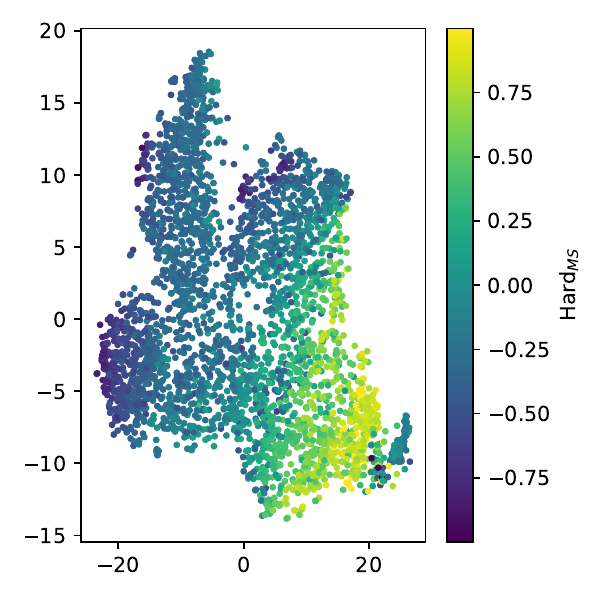}
        \caption{\texttt{hard\_ms}}
        \label{fig:hard_ms}
    \end{subfigure}
    }
    \caption{\ac{tSNE} visualization color-coded using three hardness ratios. Colors range from blue (lower hardness ratios) to yellow (higher hardness ratios). Similar hardness ratio values tend to be close to each other.}
    \label{fig:tsnehardness}
\end{figure*}

\autoref{fig:hardness_slice} compares mean reconstructed and ground-truth spectra (similarly to \autoref{fig:smoothed_collective}) for three ranges of \texttt{hard\_hs} (see \autoref{tab:tabulated_spectra}): $[-1, -0.33]$, $[-0.33, 0.33]$ and $[0.33, 1]$ across the full test set, corresponding to soft, intermediate, and hard X-ray spectra, respectively. The hardness ratio \texttt{hard\_hs} $= (F_\mathrm{hard} - F_\mathrm{soft})/(F_\mathrm{hard} + F_\mathrm{soft})$ quantifies the relative balance between the soft band ($\sim 0.5$--$1.2$\,keV) and the hard band ($\sim 2$--$8$\,keV) in Chandra data. The reconstruction quality varies across the energy range. For soft spectra (panel a), the model reconstructs the low-energy part of the spectrum well but shows larger discrepancies at higher energies. This behaviour is consistent with regression to the mean, a well-known tendency of neural networks to predict values close to the dataset average for inputs at the extremes of the training distribution. Sources with very soft spectra provide the network with very few hard-band photons, so predictions revert toward mean values in that energy regime. Importantly, however, the $1\sigma$ deviations of the predictions still encompass the range of observed hard photons. For hard spectra (panel c), the reconstruction is more uniform across energy ranges. A limitation in spectral reconstruction at extreme energies does not imply that the latent representation fails to encode the corresponding spectral information. As \autoref{fig:tsnehardness} and \autoref{tab:all_regression} show, the latent embeddings correlate strongly with the hardness ratios: for \texttt{hard\_hs}, the mean absolute error is $0.18$, representing a $64\%$ improvement over the mean baseline, with an $R^2$ of $0.83$. This indicates that the latent space encodes the global balance between soft and hard X-ray emission, even when the pixel-level reconstruction of the hard band is imperfect for the softest sources.

As expected from the contingency matrix, spectra from different classes and clusters exhibit some shared features. For instance, \acp{HM-STAR}, \acp{LM-STAR}, and \acp{YSO} spectra tend to be softer. This does not prevent the latent space from encoding physical properties, which can be used for downstream regression, as we show in \autoref{sec:results_regression}. In general, the latent space is more informative of the summary statistics, such as the hardness ratios, than it is of the specific classes of objects.

To conclude, this clustering analysis demonstrates that the autoencoder can capture continuous physical properties better than discrete astrophysical class labels (see also \autoref{sec:physical_correlations}). The overlap of source types within the latent space is not a failure of feature extraction, but a consequence of inter-class similarity. This analysis confirms that the learned representations are physically grounded, even if the spectra of sources with higher hardness ratios are better modeled because of their abundance.

\subsubsection{Dimensionality reduction}

\begin{figure*}[ht!]
\centering
\begin{subfigure}[b]{0.58\textwidth}
    \centering
    \includegraphics[width=\textwidth]{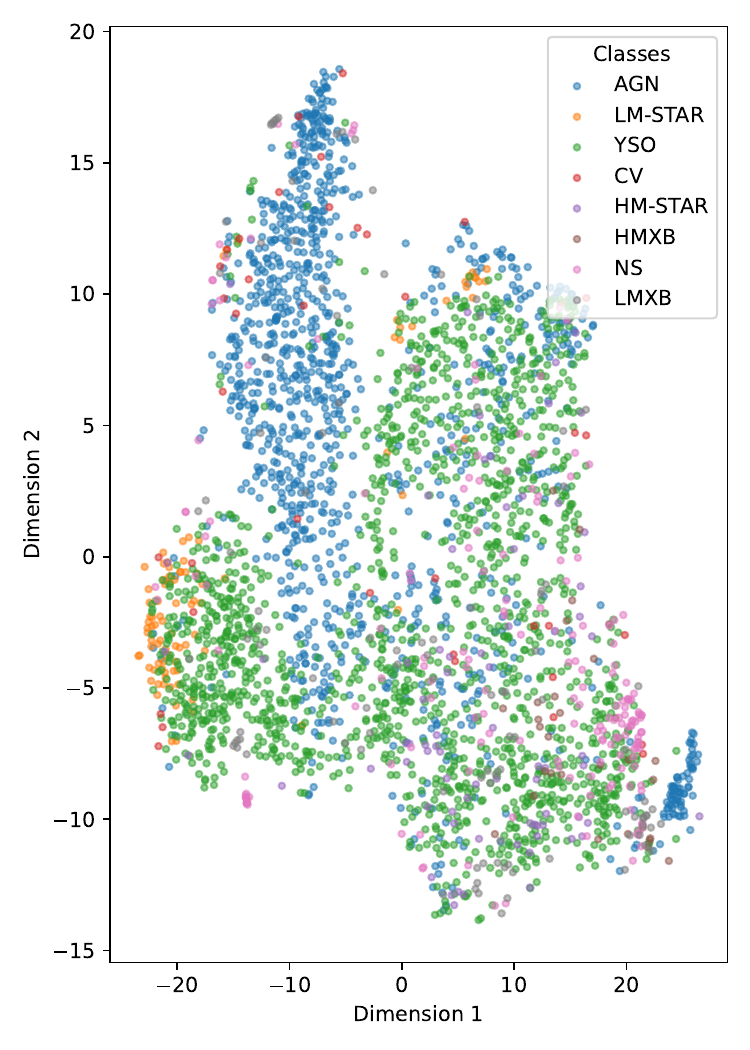}
    \caption{\ac{tSNE} for all the classes}
    \label{fig:tsne_total}
\end{subfigure}%
\hfill
\begin{minipage}[b]{0.39\textwidth}
    \centering
    \begin{subfigure}[b]{\textwidth}
        \centering
        \includegraphics[width=\textwidth]{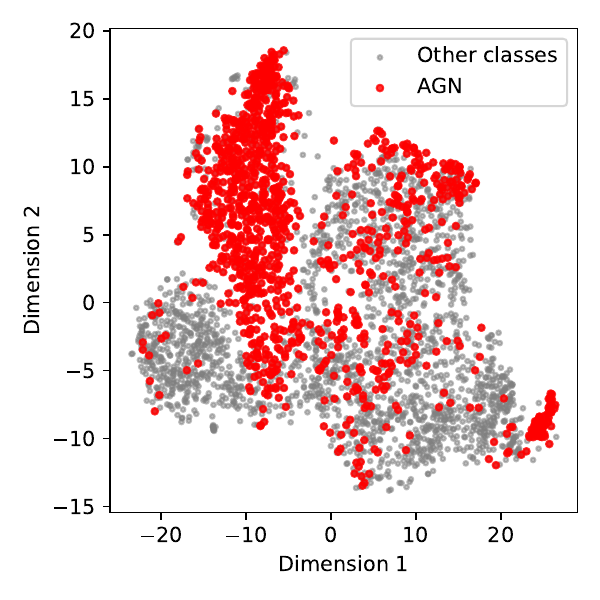}
        \caption{\ac{tSNE} for \acp{AGN}}
        \label{fig:tsne_AGN}
    \end{subfigure}

    \begin{subfigure}[b]{\textwidth}
        \centering
        \includegraphics[width=\textwidth]{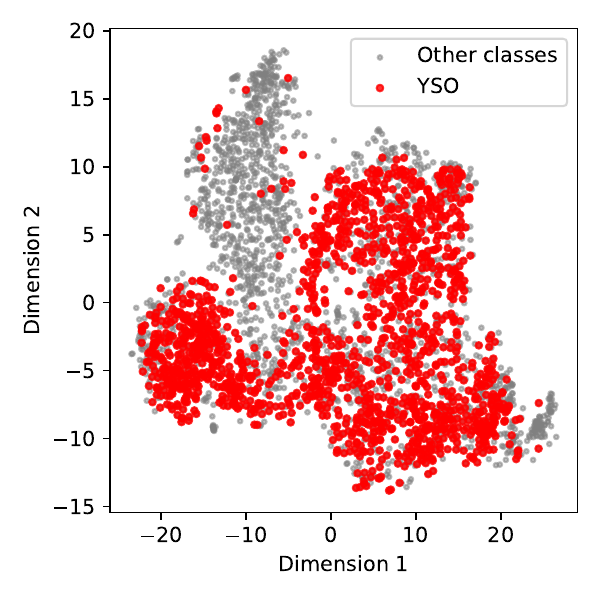}
        \caption{\ac{tSNE} for \acp{YSO}}
        \label{fig:tsne_YSO}
    \end{subfigure}
\end{minipage}
\caption{2-dimensional \ac{tSNE} representations of the autoencoder latent space for (a) all the test set objects with assigned classes, (b) \acp{AGN} examples, and (c) \acp{YSO} examples.}
\label{fig:tsne_all}
\end{figure*}

Figure \ref{fig:tsne_all} shows a 2-dimensional representation of the autoencoder latent space after training, as obtained using the \ac{tSNE} dimensionality reduction method. 

Although the autoencoder was not trained for classification, the latent space can be used to differentiate between some classes, while acknowledging that some confusion remains in underrepresented classes. This highlights the importance of the self-supervised pre-training of the autoencoder and its potential as a foundation model for X-ray datasets. Section \ref{sec:results_regression}, however, shows that the autoencoder is effective at regression tasks, as it captures spectral properties.  
The most represented classes (see also \autoref{fig:tsne_all}) are \ac{AGN} (blue) and \ac{YSO} (green), with the latter partially overlapping in \ac{tSNE} space with less-represented classes, such as \acp{LM-STAR} (orange) and \acp{NS} (pink).  Considering the case of \acp{LM-STAR}, their mean spectra appear visually different from those of \acp{YSO}, but their 1$\sigma$ standard deviation bands overlap significantly across most energy bins, as reflected by m < 1 for $\approx$ 78\% of bins in the predictions and $\approx 86\%$ in the \ac{GT} (see also \autoref{sec:degeneracies}).

The two predominant classes, on the other hand, occupy  distinct regions in the learned representation: \acp{YSO} are concentrated mostly in the bottom area, and most \acp{AGN} are clustered in the upper-left part of the plot. \acp{AGN} span a wider range due to their physical heterogeneity and their spectral shapes partially overlapping with those of some \acp{YSO} (see also \autoref{fig:smoothed_collective}). The overlaps with other classes depend on the intra-class variability and inter-class similarity of spectra, that for some couples of classes are not significantly different, as shown in \autoref{sec:degeneracies}.

Despite evident overlap between the \acp{YSO} and other types, some classes occupy adjacent, relatively distinct regions that indicate similarities in the X-ray spectral shape, which can be attributed to specific emission mechanisms. For example, the X-ray luminosities and overall spectral features of M-dwarf stars and \acp{YSO} can be similar, as they are both powered by magnetic activity near their surfaces. While both groups exhibit similar spectral features driven by magnetic activity, their separation suggests distinct activity regimes (e.g., accretion-dominated vs. magnetically-dominated) rather than fundamentally different spectral states. %

Interestingly, \ac{tSNE} reveals a distinct cluster of \acp{AGN} in the bottom right corner of the bidimensional representation of the latent space, near \acp{YSO} and \acp{NS}, as they have similar physical properties (e.g., \texttt{hard\_hm} and \texttt{hard\_hs}, see also \autoref{fig:tsnehardness}). Upon closer inspection, all observations within this region correspond to multiple distinct detections of a single source: 2CXO J195928.3+404401. This source stands out due to its significantly higher hardness ratios compared to the rest of the \ac{AGN} population in the dataset. 2CXO J195928.3+404401 is the nucleus of Cygnus A \citep{carilli_cygnus_1996}, a powerful radio galaxy with strong jets coming out of the \ac{SMBH}, and shocking the surrounding medium, which results in a very hard spectrum, compared with both other radio galaxies and average \acp{AGN}. Moreover, that the spectrum of this source is among the hardest in the dataset, and it is well distinct from the bulk of the \ac{AGN} class (see \autoref{fig:tsnehardness}). %

\begin{figure}
    \centering
    \includegraphics[width=0.8\linewidth]{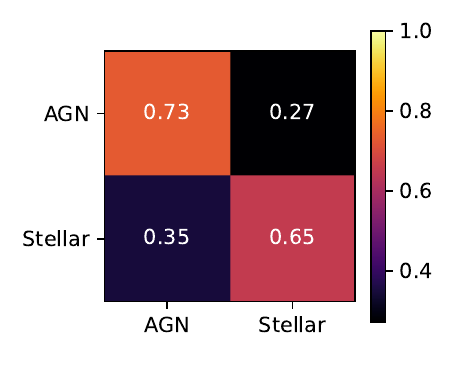}
    \caption{The contingency matrix for \acp{AGN} and stellar-mass compact objects, with a balanced accuracy of $\approx 69\%$.}
    \label{fig:cm_agn_xb}
\end{figure}

\begin{figure}
    \centering
    \includegraphics[width=\linewidth]{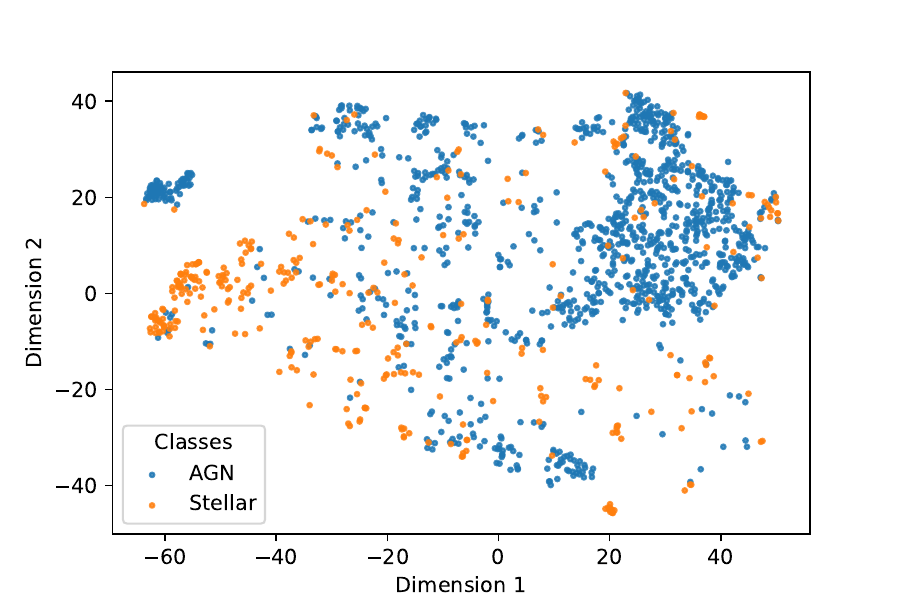}
    \caption{The output of \ac{tSNE} for clustering \acp{AGN} (blue) and stellar-mass compact objects (orange).}
    \label{fig:tsne_agn_xb}
\end{figure}

The \ac{tSNE} maps suggest that the representations that the autoencoder has learned are useful for classification. We now investigate to what extent the learned latent embeddings can discriminate between \acp{AGN} and stellar-mass compact objects. This is a relevant task because, in the absence of a redshift estimation, these two types of accretion-powered systems can be spectrally similar, despite the difference in the mass of the central accretor. In particular, the spectral photon index ($\Gamma$), which parametrizes the slope of a power law spectrum, for X-ray binaries in the low/hard state has a similar range compared to AGNs ($\Gamma \sim 1.5$) and similarities have been reported in the relationship between the X-ray spectral hardness and the Eddington ratio in both \acp{AGN} and stellar-mass compact objects \citep{trichas_empirical_2013}. We group stellar-mass compact objects to include \acp{LMXB}, \acp{HMXB}, \acp{NS}, and \acp{CV}, and compare their latent representations to objects classified as \acp{AGN}. We fit \ac{GMM} with the best hyperparameters found for the 8-class clustering on the reduced test set containing only objects from the two selected classes. Figure \ref{fig:cm_agn_xb} shows the contingency matrix, corresponding to a balanced accuracy of $\approx 69\%$. Figure \ref{fig:tsne_agn_xb} presents the output of \ac{tSNE} on the latent space for the reduced test set. We find that spectral information alone is sufficient to partially separate \acp{AGN} for compact objects of stellar mass, with the \acp{AGN} occupying the upper right corner of the representation space. As we will see in \autoref{sec:results_regression}, this relates to the average hardness ratio of the sources, with most \acp{AGN} having intermediate \texttt{hard\_hs} values, and most stellar-mass compact objects displaying a relatively high hardness ratio in the same energy band, with the exception of the outlying Cygnus A detections, which are closer in the representation space to \acp{LMXB} and close to each other. %

\subsubsection{Comparative performance}

Overall, the representations learned using the autoencoder do not underperform with respect to other automatic classification methods that only use the X-ray data. For example, \citet{perez-diaz_unsupervised_2024} achieve a 61\% accuracy in a 4-type classification task (relying on \ac{GMM}) using tabulated properties from the \ac{CSC} (v 2.0), which do not have the same representation/reconstruction power as the autoencoder features learned in our method. However, methods that use the full information from the event files in a representation learning framework, rather than only the spectral information, might have an advantage over spectral-only methods like the one presented here, as they can use variability information as well. \citet{song_poisson_2025} achieve a 60\% accuracy in a similar 8-label task. However, their approach differs fundamentally, as they use a neural field estimator to model the photon arrival times (light curves), capturing variability information which are not included in spectra. In contrast, our method relies solely on spectral shapes. Moreover, the analysis of \citet{song_poisson_2025} is more computationally expensive at inference time because an optimization loop (gradient descent) is used to find the latent vector that best fits the data.

\subsection{Physical correlations}
\label{sec:physical_correlations}

\begin{figure*}[htbp]
    \centering
    \begin{subfigure}[b]{0.24\textwidth}
        \centering
        \includegraphics[width=\textwidth]{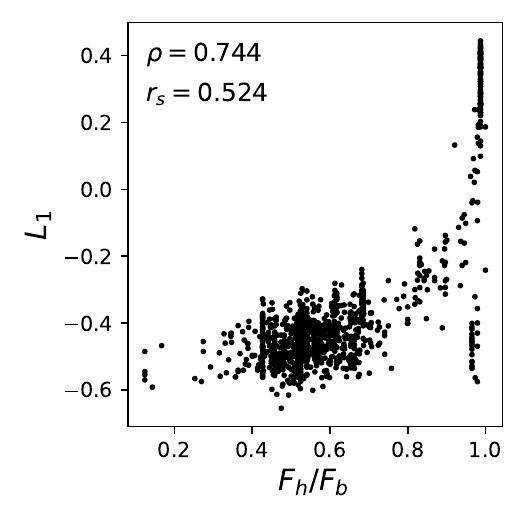}
        \caption{$L_1$}
    \end{subfigure}\hfill
    \begin{subfigure}[b]{0.24\textwidth}
        \centering
        \includegraphics[width=\textwidth]{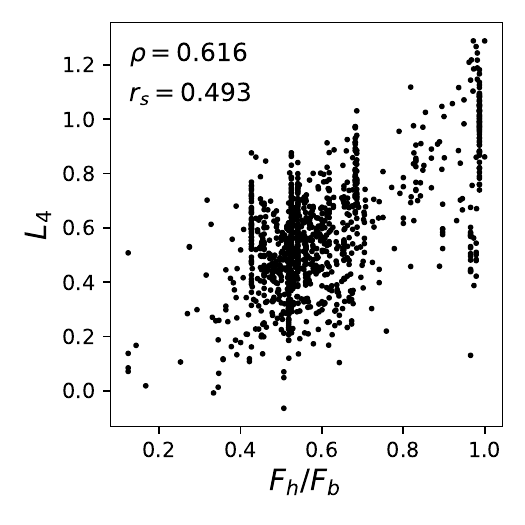}
        \caption{$L_4$}
    \end{subfigure}\hfill
    \begin{subfigure}[b]{0.24\textwidth}
        \centering
        \includegraphics[width=\textwidth]{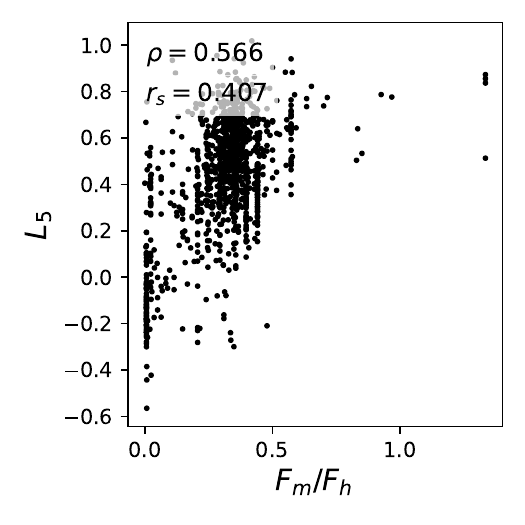}
        \caption{$L_5$}
    \end{subfigure}\hfill
    \begin{subfigure}[b]{0.24\textwidth}
        \centering
        \includegraphics[width=\textwidth]{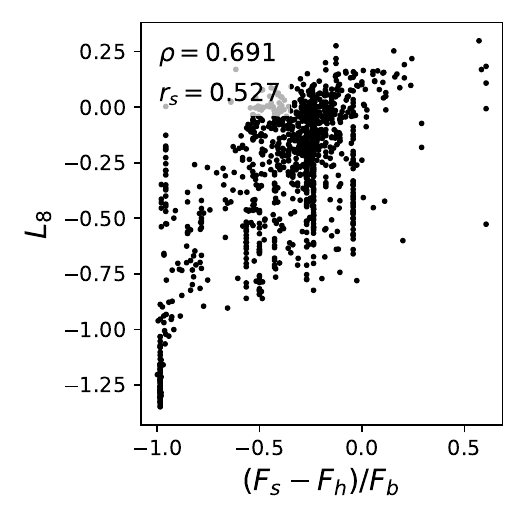}
        \caption{$L_8$}
    \end{subfigure}
    \caption{Relevant correlations between latent space dimensions and physical formulas obtained from tabulated data for \acp{AGN}.}
    \label{fig:latent_physical}
\end{figure*}

Features learned using self-supervised approaches such as the one used here, either compare fairly or over perform with respect to methods that use manually extracted features in downstream tasks of classification. For example, as we have pointed out, our spectral classifier has a similar accuracy to the one presented in \citet{perez-diaz_unsupervised_2024}, which uses very carefully crafted, human-interpretable features manually extracted as part of the \ac{CSC} processing, which include both spectral and time variability properties. Also, \citet{song_poisson_2025} demonstrate that automatically learned latent features for light curve reconstruction lead to an F1 score of 0.69 in the YSO vs. AGN classification task, comparable to algorithms that use a much richer set of multi-wavelength features \citep[e.g.,][]{yang_classifying_2022}. 

However, it is of little use having specialized automatically extracted features if they are not interpretable. If they overperform with respect to manually extracted features at classification, it is reasonable to expect that they relate to physically interpretable features, or to some combination of physical features that provide them with the observed predicted power. In order to investigate this, we performed symbolic regression to mathematically relate the latent features ${L_i}$ to physically interpretable quantities, such as X-ray fluxes and hardness ratios. In the following paragraphs, we focus on sources of classes \ac{AGN} and \ac{YSO}, as the majority of astrophysical sources at high energies falls in these two groups. 

Figure \ref{fig:latent_physical} shows some of the most significant correlations for \acp{AGN} and reports both the Pearson correlation coefficient $\rho$ and the Spearman correlation coefficient $r_s$ for each case. Latent features $L_1$ and $L_4$ are respectively strongly and moderately correlated with the fractional flux emitted in the hard X-ray band (2--8\,keV), $F_h/F_b$, a crucial diagnostic of the physical nature of the emission, as non-thermal processes tend to dominate emission in the hard band. \acp{AGN} with high $L_1$ and $L_4$ values are therefore likely to emit strongly through non-thermal processes such as inverse Comptonization, or they can also be heavily obscured AGNs, for which the soft X-ray emission is absorbed by dust. Latent feature $L_5$ is moderately correlated ($\rho \approx 0.566$, $r_s \approx 0.407$) with $F_m/F_h$, the ratio between the medium band (1.2--2\,keV) flux and the hard band flux, which provides additional information about the emission mechanisms. Sources with enhanced medium band emission include thermal plasma sources with $kT \leq 1 - 2$~keV, such as supernova remnants and coronal emission from active stars, as well as sources with strong atomic emission lines in the 1--2~keV range, such as X-ray binaries with photoionized winds. Finally, the eighth latent dimension ($L_8$) is moderately correlated ($\rho \approx 0.691$, $r_s \approx 0.527$) with $(F_s - F_h)/F_b$, the relative fractional emission in the soft band (0.5--1.2\,keV) with respect to the hard band. Sources with a high $L_8$ are thus likely to include super soft sources AGNs, which are dominated by the thermal emission from the accretion disk, showing no hard coronal emission. Tidal Disruption Events also populate this soft regime.

\begin{figure*}[htbp]
    \centering
    \begin{subfigure}[b]{0.24\textwidth}
        \centering
        \includegraphics[width=\textwidth]{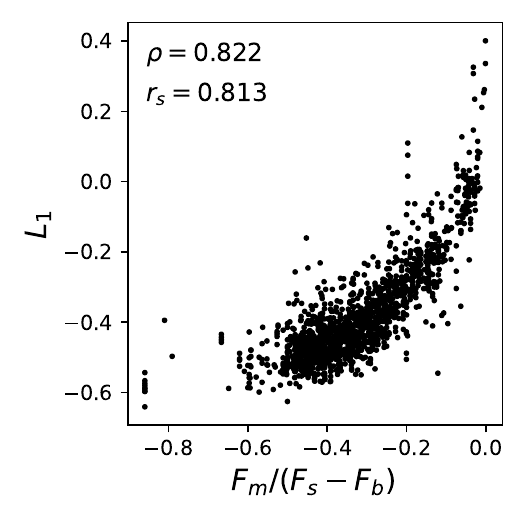}
        \caption{$L_1$}
    \end{subfigure}\hfill
    \begin{subfigure}[b]{0.24\textwidth}
        \centering
        \includegraphics[width=\textwidth]{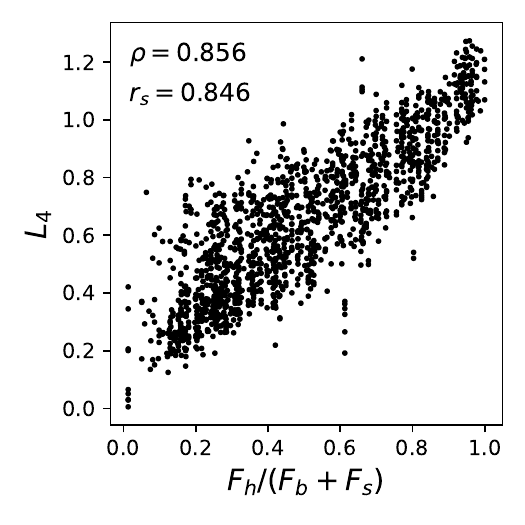}
        \caption{$L_4$}
    \end{subfigure}\hfill
    \begin{subfigure}[b]{0.24\textwidth}
        \centering
        \includegraphics[width=\textwidth]{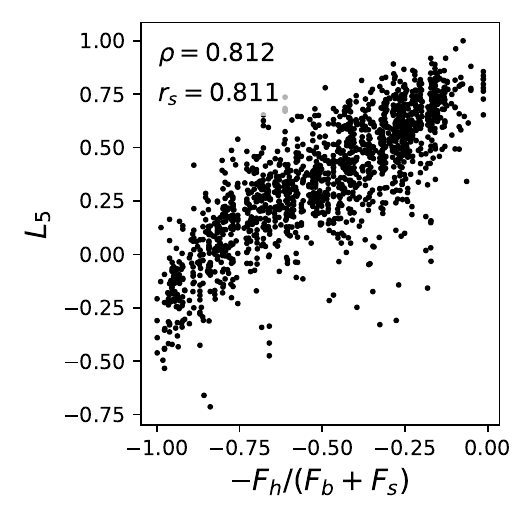}
        \caption{$L_5$}
    \end{subfigure}\hfill
    \begin{subfigure}[b]{0.24\textwidth}
        \centering
        \includegraphics[width=\textwidth]{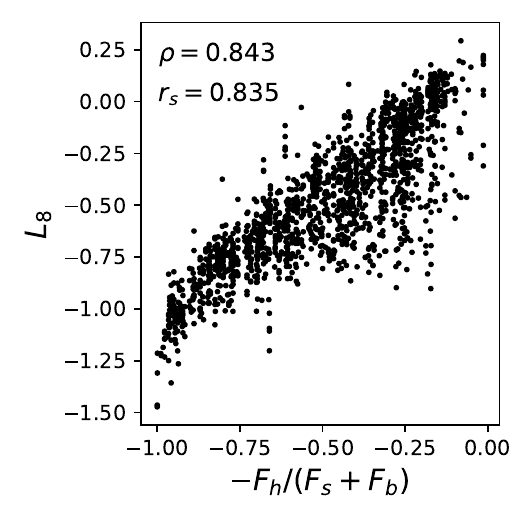}
        \caption{$L_8$}
    \end{subfigure}
    \caption{Relevant correlations between latent space dimensions and physical formulas obtained from tabulated data for \acp{YSO}.}
    \label{fig:latent_physical_yso}
\end{figure*}

\autoref{fig:latent_physical_yso} shows the derived  correlations for \acp{YSO}. Latent features $L_4$ and $L_5$ are strongly correlated/anticorrelated  with $F_h/(F_b+F_s)$, which relates to the fractional flux emitted in the hard X-ray band. Hard X-ray emission in \acp{YSO} is associated to coronal particle acceleration by magnetic reconnection events that produce X-ray flares, which are more common in young, rapidly rotating stars. Therefore, latent features $L_4$ and $L_5$ are good indicators of magnetic activity in stars. Latent feature $L_1$ is strongly anticorrelated with $F_m/(F_b-F_s)$, which is related to the fraction of the X-ray flux emitted in the 1-2~keV range. So, a low value of $L_1$ might indicate sources with thermal coronal plasma at intermediate temperatures, $kT \sim 0.8-1.5$~keV ($\sim 10-20 \times 10^6$~K ), usually peaking in this energy range. 

These automatically learned features thus represent spectral summaries that have a direct link to specific physical mechanisms of X-ray emission in both \acp{AGN} and \acp{YSO}. They are an alternative to and are related to commonly used diagnostics such as the hardness ratios, but carry more information than the latter about the specific spectral shape and time domain properties of X-ray sources, and significantly save computational costs associated with computing these human-extracted properties.

\subsection{Regression}
\label{sec:results_regression}

We now investigate if the representations relate to human-interpretable properties that are associated with the physics of the X-ray sources. Specifically, we train a Huber regressor to predict the physically interpretable properties described in  Table \ref{tab:tabulated_spectra} using the learned representations as input. 

Table \ref{tab:all_regression} reports the results for the entire test set (column "All test set") as well as for each variable computed only on the subset of samples corresponding to the best-fitting spectral model. The "best model" refers to the spectral model that achieved the best fit statistic for each source. For example, for \verb%powlaw_gamma%, in the "best model" set, we only consider sources that were best fitted with the power-law model. All results are compared against a baseline  (in which the mean value of each target variable is used as the predicted value for the entire test set). The column "Impr. (\%)" indicates the percentage improvement with respect to the mean baseline. To illustrate the usefulness of the representations for regression on a crucial summary statistic, namely the hardness ratio, in Figure \ref{fig:tsnehardness} we show the correlation between the learned latent representations and this quantity using a color-coded \ac{tSNE} projection. We note a clear correlation between the latent space and the three hardness ratios, which describe the overall spectral shape of the X-ray sources.

Apart from the hardness ratios, which are model-independent quantities well captured by the learned neural summaries, \autoref{tab:all_regression} also lists several model-dependent physical parameters, such as the power law spectral slope ($\Gamma$) which indicates the acceleration of electrons in the hot plasma via synchrotron emission or inverse Comptonization, the blackbody temperature (\verb|bb_kt|) when a thermal component is present, the hydrogen column density in the line of sight to the source (\verb|n_H|), and even non-spectral diagnostics, such as the time variability of the X-ray flux (\verb|var_prob_b|) of the source. When appropriate spectral models are considered ("Best model set" in the table), information gain with respect to the baseline model can be well above 50\% for the hydrogen column densities, and close to 50\% for the black body temperatures. There is information gain also for parameters such as temporal flux variability, which is note intuitively related to the spectral shape of the source. This indicates that the features learned through spectral reconstruction capture multiple information about X-ray sources, and that there is mutual information between their spectral and temporal properties.

We also compare the performance of our approach to other self-supervised learning methods, such as the Poisson Process Autodecoder (PPAD) presented in \cite{song_poisson_2025}. The latter is a ResNet architecture trained to predict the mean Poisson arrival rate of individual X-ray photons in a particular energy range, using the individual photon event recordings (as opposed to the spectrum) as the input. Table \ref{tab:performance_comparison_song} shows the results. The PPAD performs better in regression tasks with respect to our model, which indicates the advantage of using the fine-grained individual photon information with respect to coarser spectral bins. We note, however, that event lists can be extremely long (tens of thousands of photons), and that the PPAD is inherently inefficient at inference time, which makes it an unlikely tool for the analysis of large repositories of X-ray sources.

\begin{table*}[ht]
\centering
\caption{Regression results. "Model" refers to the \ac{MAE} of our method, while "Base" refers to the baseline. For global variables (top), model-specific sets are not applicable.}
\label{tab:all_regression}
\setlength{\tabcolsep}{4pt} %

\begin{tabular}{@{}l ccc cccc@{}}
\toprule
 & \multicolumn{3}{c}{\textbf{All test set}} & \multicolumn{4}{c}{\textbf{Best model set}} \\ 
 \cmidrule(lr){2-4} \cmidrule(l){5-8} 
\textbf{Variable} & \textbf{Model} & \textbf{Base} & \textbf{Impr.\%} & \textbf{Model} & \textbf{Base} & \textbf{Impr.\%} & \textbf{Samples} \\ \midrule
hard\_hs      & 0.18 & 0.49 & 64 & \textemdash & \textemdash & \textemdash & \textemdash \\
hard\_ms      & 0.14 & 0.35 & 60 & \textemdash & \textemdash & \textemdash & \textemdash \\
hard\_hm      & 0.11 & 0.27 & 60 & \textemdash & \textemdash & \textemdash & \textemdash \\
var\_prob\_b  & 0.35 & 0.40 & 13 & \textemdash & \textemdash & \textemdash & \textemdash \\
var\_index\_b & 3.12 & 3.69 & 15 & \textemdash & \textemdash & \textemdash & \textemdash \\ \addlinespace

powlaw\_gamma & 0.62 & 0.90 & 32 & 0.56 & 0.86 & 35 & 1609 \\
powlaw\_nh    & 49.96 & 95.46 & 48 & 47.85 & 85.89 & 44 & 1609 \\
bb\_kt        & 1.29 & 2.38 & 46 & 1.01 & 1.87 & 46 & 188 \\
bb\_nh        & 34.70 & 70.09 & 50 & 144.37 & 289.97 & 50 & 188 \\
brems\_kt     & 9.26 & 15.65 & 41 & 3.31 & 4.70 & 30 & 541 \\
brems\_nh     & 74.46 & 155.04 & 52 & 33.74 & 77.28 & 56 & 541 \\
apec\_kt      & 2.59 & 3.72 & 30 & 1.03 & 1.37 & 25 & 904 \\
apec\_abund   & 0.41 & 0.49 & 17 & 0.33 & 0.41 & 20 & 904 \\
apec\_z       & 0.13 & 0.15 & 8 & 0.06 & 0.07 & 10 & 904 \\
apec\_nh      & 50.22 & 100.42 & 50 & 23.02 & 53.95 & 57 & 904 \\ \bottomrule
\end{tabular}
\end{table*}

\begin{table*}[ht]
\centering
\begin{tabular}{lcccc}
\toprule
\textbf{Variable} & \multicolumn{2}{c}{\textbf{MSE}} & \multicolumn{2}{c}{\textbf{R$^2$}} \\
\cmidrule(r){2-3} \cmidrule(l){4-5}
                  & This work & \cite{song_poisson_2025} & This work & \cite{song_poisson_2025} \\
\midrule
hard\_hs & 0.06 & 0.01 & 0.83 & 0.94 \\
hard\_ms & 0.04 & 0.02 & 0.77 & 0.87 \\
hard\_hm & 0.02 & 0.01 & 0.82 & 0.88 \\
\bottomrule
\end{tabular}
\caption{Comparison of hardness ratio estimations with the work in \cite{song_poisson_2025}.}
\label{tab:performance_comparison_song}
\end{table*}

\section{Summary and conclusions}
\label{sec:summary}

In this work, we addressed the challenges of characterizing large catalogs of high-energy X-ray sources. To this end, we presented a self-supervised deep learning framework based on a transformer autoencoder that generates compressed, low-dimensional representations of Chandra X-ray spectra. Our primary objective is to evaluate whether these learned latent representations can capture fundamental physical properties and serve as a robust alternative to manually extracted features for downstream tasks, including source classification, parameter regression, and symbolic regression, at lower computational cost.

The autoencoder compresses the spectra into an 8-dimensional latent space and preserves enough physical information about each source to allow for spectral reconstruction. The latent space has clustering properties that generally preserve astrophysical classes despite not having been trained on clustering. Known degeneracies in spectral shape between classes are also preserved, although the model is able to successfully separate between large accretors, mostly AGNs, and stellar-size accretors (e.g., X-ray binaries) when redshift information is not available. When clustering observations into 8 classes (\ac{AGN}, \ac{CV}, \ac{HM-STAR}, \ac{HMXB}, \ac{LM-STAR}, \ac{LMXB}, \ac{NS}, or \ac{YSO}) the balanced accuracy is $\sim$40\%, but it significantly increases for binary classification between \acp{AGN} and stellar-mass compact objects, reaching $\sim$69\%, comparable to the performance of methods that use human-extracted features. The learned representation also preserve summaries that are relevant to the identification of astrophysically compelling sources that are being actively sought in large X-ray catalogs, such as TDEs, QPEs, and FXTs, including summaries related to temporal variability. Our framework is therefore interpretable and immediately applicable to dedicated searches for these objects. As next-generation X-ray surveys continue to produce vast amounts of spectral data, unsupervised and self-supervised methods will be able to capture relevant information, which in turn can be used for classification, discovery, and the training of foundation models in astrophysics.

Our approach has some limitations. The test set consists of $\sim$3,200 labeled spectra, and class imbalance impacts clustering performance. Some classes, such as \acp{LMXB}, are underrepresented, and some are spectrally similar to other types, making them more challenging to distinguish. Moreover, this work does not consider uncertainties on fluxes nor faint sources. These constraints highlight the need for more comprehensive labelled datasets and techniques to incorporate uncertainty in reconstruction and self-supervised learning.

Improvements to our proposed framework can incorporate uncertainties \citep{farrell_deducing_2023} in the flux values to provide uncertainty estimations for the predictions. 
The regression techniques presented can also be used to estimate physical quantities from the latent space. The method can also be adapted to incorporate observations across multiple wavelengths and data in multiple modalities \citep{cuoco_computational_2022, pinciroli_vago_deepgravilens_2023, buck_deep_2024, rizhko_self-supervised_2024, casado_multimodal_2024}. 
A deeper analysis of the similarities between the variability in spectra and light curves can also be an area for further research.

This work contributes to the broader field of representation learning in astrophysics by demonstrating its applicability to high-energy spectral data. Future extensions will include multimodal learning by combining spectra with temporal or spatial information, anomaly detection in the latent space to discover rare sources, and transfer learning across different X-ray instruments.

\begin{acknowledgements}
NOPV acknowledges support from INAF and CINECA for granting 125,000 core hours on the Leonardo supercomputer (project INA24\_C6B05). RA acknowledges financial support from INAF through the grant ``INAF-Astronomy Fellowships in Italy 2022 - (GOG)'' and the Bando Ricerca Fondamentale INAF 2024 Large Grant ``Timing the Ultra Luminous x-ray Pulsars (TULiP)''. The authors gratefully acknowledge the AstroAI multimodal group at CfA for their support, valuable suggestions and insightful discussions. This research has made use of data obtained from the Chandra Source Catalog, provided by the Chandra X-ray Center (CXC). This research has also made use of software provided by the Chandra X-ray Center (CXC) in the application packages CIAO\footnote{\url{https://cxc.cfa.harvard.edu/ciao/}} and Sherpa\footnote{\url{https://cxc.cfa.harvard.edu/sherpa/}}.
\end{acknowledgements}

\bibliographystyle{aa} %
\bibliography{zotero_biblio} %

\begin{acronym}
\acro{AD}{anomaly detection}
\acro{AGN}{active galactic nucleus}
\acro{AI}{artificial intelligence}
\acro{ANN}{artificial neural network}
\acro{APEC}{astrophysical plasma emission code}
\acro{BIRCH}{Balanced Iterative Reducing and Clustering using Hierarchies}
\acro{CNN}{convolutional neural network}
\acro{CSC}{Chandra Source Catalog}
\acro{CIAO}{Chandra Interactive Analysis of Observations}
\acro{CV}{cataclysmic variable}
\acro{DBSCAN}{Density-Based Spatial Clustering of Applications with Noise}
\acro{DL}{deep learning}
\acro{GMM}{gaussian mixture model}
\acro{GP}{genetic programming}
\acro{GT}{ground truth}
\acro{HM-STAR}{high-mass star}
\acro{HMXB}{high-mass X-ray binary}
\acro{HSC}{Hyper Suprime-Cam}
\acro{IQR}{interquartile range}
\acro{LinR}{linear regression}
\acro{LM-STAR}{low-mass star}
\acro{LMXB}{low-mass X-ray binary}
\acro{LR}{learning rate}
\acro{MAE}{mean absolute error}
\acro{ML}{machine learning}
\acro{MSE}{mean square error}
\acro{NS}{pulsar or isolated neutron star}
\acro{OPTICS}{Ordering Points To Identify Cluster Structure}
\acro{PCA}{principal component analysis}
\acro{PHA}{pulse height amplitude}
\acro{PI}{pulse invariant}
\acro{PSF}{point spread function}
\acro{SNR}{signal to noise ratio}
\acro{SR}{symbolic regression}
\acro{tSNE}{t-distributed stochastic neighbor embedding}
\acro{RMF}{redistribution matrix file}
\acro{ARF}{auxiliary response file}
\acro{XB}{X-ray binary}
\acro{WGAN}{Wasserstein generative adversarial network}
\acro{YSO}{young stellar object}
\acroplural{AGN}[AGNs]{active galactic nuclei}
\acroplural{HMXB}[HMXBs]{high-mass X-ray binaries}
\acroplural{LMXB}[LMXBs]{low-mass X-ray binaries}
\acroplural{XB}[XBs]{X-ray binaries}
\acro{SMBH}{supermassive black hole}
\end{acronym}

\begin{appendix}
\onecolumn
\section{Physical variables}
\begin{table}[htbp]
\centering
\caption{Physical variables collected for showing their correlation with the latent space extracted by the autoencoder.}
\label{tab:tabulated_spectra}
\begin{tabular}{p{0.15\linewidth} p{0.75\linewidth}}
    \toprule
    \textbf{Variable} & \textbf{Description} \\ 
    \midrule
$F_s$ & Average soft band flux (0.5 - 1.2 keV) \\ 
$F_b$ & Average broad band flux (0.5 - 7 keV) \\ 
$F_m$ & Average medium band flux (1.2 - 2 keV) \\ 
$F_h$ & Average hard band flux (2 - 7 keV) \\ 
hard\_hs & hard (2.0-7.0 keV) - soft (0.5-1.2 keV) energy band hardness ratio \\ 
hard\_ms & medium (1.2-2.0 keV) - soft (0.5-1.2 keV) energy band hardness ratio \\ 
hard\_hm & hard (2.0-7.0 keV) - medium (1.2-2.0 keV) energy band hardness ratio \\ 
var\_prob\_b & Gregory-Loredo variability probability values \\ 
var\_index\_b & Intra-observation Gregory-Loredo variability index \\ 
powlaw\_gamma & photon index, defined as $F_E \propto E^{-\gamma}$, of the best fitting absorbed power-law model spectrum to the source region aperture \ac{PI} spectrum \\ 
powlaw\_nh & $N_H$ column density of the best fitting absorbed power-law model spectrum to the source region aperture \ac{PI} spectrum \\ 
bb\_kt & temperature (kT) of the best fitting absorbed black body model spectrum to the source region aperture \ac{PI} spectrum \\ 
bb\_nh & $N_H$ column density of the best fitting absorbed black body model spectrum to the source region aperture \ac{PI} spectrum \\ 
brems\_kt & temperature (kT) of the best fitting absorbed bremsstrahlung model spectrum to the source region aperture \ac{PI} spectrum \\ 
brems\_nh & $N_H$ column density of the best fitting absorbed bremsstrahlung model spectrum to the source region aperture \ac{PI} spectrum \\ 
apec\_kt & temperature (kT) of the best fitting absorbed \ac{APEC} model spectrum to the source region aperture \ac{PI} spectrum \\ 
apec\_abund & abundance of the best fitting absorbed \ac{APEC} model spectrum to the source region aperture \ac{PI} spectrum \\ 
apec\_z & redshift of the best fitting absorbed \ac{APEC} model spectrum to the source region aperture \ac{PI} spectrum \\ 
apec\_nh & $N_H$ column density of the best fitting absorbed \ac{APEC} model spectrum to the source region aperture \ac{PI} spectrum \\ \bottomrule
\end{tabular}%
\end{table}

\twocolumn
\section{Methods}
\subsection{Autoencoder}
\label{app:method_autoencoder}
Autoencoders consist of two main components \citep{yang_autoencoder-based_2022}: the encoder and the decoder. The encoder maps the input data (the resampled normalized spectra $\hat f$) into a lower-dimensional space known as the latent space. Deep encoders are typically implemented using multiple layers of \acp{ANN} \citep{hinton_reducing_2006, wang_auto-encoder_2016}, such as fully connected networks \citep{bengio_representation_2013}, \acp{CNN} \citep{masci_stacked_2011}, or recurrent architectures \citep{sutskever_sequence_2014}, which progressively reduce the dimensionality of the input. For a single training example, the output of the encoder is a low-dimensional latent vector, denoted as $z$. The set of latent vectors for all examples captures the salient characteristics of the input data while discarding irrelevant information.

The decoder, on the other hand, takes as input the latent vectors and attempts to reconstruct the original input data $\hat f$ as $\check f$. Like deep encoders, deep decoders consist of multiple layers of a neural network.  The quality of the reconstruction, as measured by an appropriate loss function, such as the \ac{MAE}, indicates how effectively the latent vectors capture the critical information in the input data. Accurate reconstruction demonstrates that the encoder has successfully preserved the most significant features in the latent space.

Autoencoders are trained to minimize the reconstruction error, which quantifies the difference between the original input $\hat f$ and the reconstructed output $\check f$. Accordingly, the loss function $\mathcal L$ used during training depends on both $\hat f$ and $\check f$. By optimizing this loss function, the autoencoder learns to represent the input data efficiently in the latent space while retaining the ability to reconstruct it accurately.

\subsection{Clustering}
\label{app:clustering_methods}

In this work, we compare different clustering algorithms, which rely on different principles:

\begin{itemize}
    \item KMeans\footnote{\url{https://scikit-learn.org/stable/modules/generated/sklearn.cluster.KMeans.html}} \citep{steinhaus_sur_1956, macqueen_methods_1967} partitions the data into $k$ clusters, updating the cluster centroids iteratively to minimize the within-cluster variance. Each data point is assigned to the cluster with the closest centroid.
    \item \Ac{GMM}\footnote{\url{https://scikit-learn.org/stable/modules/generated/sklearn.mixture.GaussianMixture.html}} \citep{reynolds_gaussian_2009} models the data as a mixture of $k$ Gaussian distributions (or components), each with its mean, covariance, and weight. 
    \item Bayesian \ac{GMM}\footnote{\url{https://scikit-learn.org/stable/modules/generated/sklearn.mixture.BayesianGaussianMixture.html}}  \citep{rasmussen_infinite_1999, bishop_pattern_2006} extends \ac{GMM} by incorporating Bayesian priors on the distributions parameters.
    \item Agglomerative clustering\footnote{\url{https://scikit-learn.org/stable/modules/generated/sklearn.cluster.AgglomerativeClustering.html}} \citep{mullner_modern_2011, ackermann_analysis_2014} is a hierarchical clustering algorithm that starts by treating each data point as a single cluster and iteratively merges close clusters.
    \item \ac{BIRCH}\footnote{\url{https://scikit-learn.org/stable/modules/generated/sklearn.cluster.Birch.html}} \citep{zhang_birch_1996} creates a hierarchical structure with data summaries and is typically used in large datasets.
    \item Spectral clustering\footnote{\url{https://scikit-learn.org/stable/modules/generated/sklearn.cluster.SpectralClustering.html}} \citep{ng_spectral_2001} is a technique that uses the eigenvalues of a similarity matrix to reduce the dimensionality of the data before using an additional clustering algorithm (here, KMeans).
    \item \ac{DBSCAN}\footnote{\url{https://scikit-learn.org/stable/modules/generated/sklearn.cluster.DBSCAN.html}} \citep{ester_density-based_1996} does not require specifying the number of clusters. It identifies clusters based on the density of data points, grouping together points that are closely packed and marking points in low-density regions as outliers. 
    \item \ac{OPTICS}\footnote{\url{https://scikit-learn.org/stable/modules/generated/sklearn.cluster.OPTICS.html}} \citep{ankerst_optics_1999} is an extension of \ac{DBSCAN} designed to handle clusters with different densities.
\end{itemize}

The clustering algorithms hyperparameters are illustrated in detail in Table \ref{tab:summary_cluster}.

\begin{table*}
    \centering
    \caption{Overview of the selected clustering algorithms, their hyperparameters, and their clustering types. The notation \texttt{np.linspace(start, stop, num)} indicates an array of \texttt{num} evenly spaced numbers starting with \texttt{start} and ending with \texttt{stop}.}
    \label{tab:summary_cluster}
    \begin{tabular}{@{} l p{14cm} l @{}}
        \toprule
        \textbf{Algorithm} & \textbf{Hyperparameters} & \textbf{Type} \\ 
        \midrule
        \textbf{KMeans} & \texttt{init}: ['k-means++', 'random'], \texttt{max\_iter}: [30000], \texttt{tol}: [1e-5] & Hard \\ 
        \addlinespace
        \textbf{\ac{GMM}} & \texttt{covariance\_type}: ['full', 'diag', 'tied', 'spherical'], \texttt{tol}: [1e-5], \texttt{max\_iter}: [30000] & Soft \\ 
        \addlinespace
        \textbf{Bayesian \ac{GMM}} & \texttt{covariance\_type}: ['full', 'diag', 'tied', 'spherical'], \texttt{tol}: [1e-5], \texttt{max\_iter}: [30000] & Soft \\ 
        \addlinespace
        \textbf{Agglomerative} & \texttt{linkage}: ['ward', 'complete', 'average', 'single'] & Hard \\ 
        \addlinespace
        \textbf{\ac{BIRCH}} & \texttt{threshold}: \texttt{np.linspace(0.1, 0.5, 41)} & Hard \\ 
        \addlinespace
        \textbf{Spectral Clustering} & \texttt{affinity}: ['nearest\_neighbors', 'rbf'] & Hard \\ 
        \addlinespace
        \textbf{\acs{DBSCAN}} & \texttt{eps}: \texttt{np.linspace(0.1, 2, 20)}, \texttt{min\_samples}: [5, 10, 15] & Hard \\ 
        \addlinespace
        \textbf{\acs{OPTICS}} & \texttt{min\_samples}: [5, 10, 15], \texttt{xi}: [0.05, 0.1, 0.2], \texttt{min\_cluster\_size}: [0.05, 0.1, 0.2] & Hard \\ 
        \bottomrule
    \end{tabular}
\end{table*}

\subsection{\acf{tSNE}}
\label{app:tsne}

The first step of \ac{tSNE} is measuring the similarity between each pair of points in a dataset in the original high-dimensional space, combining the Euclidean distance and a Gaussian distribution. The similarity between two points $i$ and $j$ is defined as:

\begin{equation}
    p_{ij} = \frac{\exp\left(-\frac{||x_i-x_j||^2}{2\sigma_i^2}\right)}{\sum_{k \neq i} \exp\left(-\frac{||x_i-x_k||^2}{2\sigma_i^2}\right)}
\end{equation}

where $\sigma_i$ is the variance associated with the Gaussian distribution for each $x_i$. 

\ac{tSNE} representations also depend on two hyperparameters: perplexity and early exaggeration. Perplexity is related to the number of nearest neighbours considered for computing the pairwise distances to preserve. Larger perplexity values capture global structures, while smaller values emphasize local relationships. Early exaggeration is a parameter that amplifies pairwise similarities during the initial optimization stages.

In the low-dimensional space, \ac{tSNE} also defines a pairwise similarity between couples of points based on the Student's t-distribution. The algorithm aims to minimize the difference between high- and low-dimensional similarities. The optimization is performed through gradient descent.

In this work, \ac{tSNE} is applied with a perplexity of 200 and an early exaggeration of 100. All plots show points belonging to different \ac{GT} classes using different colours to facilitate comparisons.

\subsection{\Acf{SR}}
\label{app:symbolic_regression}

\begin{figure}
    \centering
    \begin{subfigure}[b]{\linewidth}
        \centering
        \includegraphics[width=0.7\linewidth]{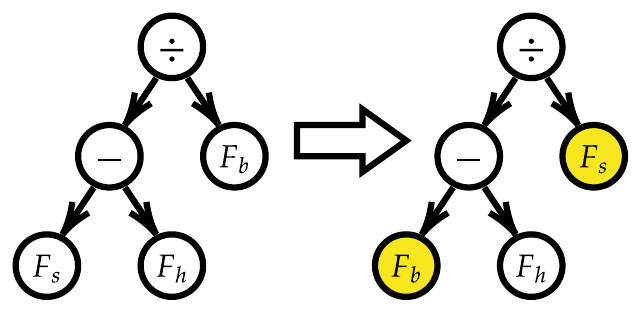}
        \caption{Mutation from the expression $(F_s - F_h)/F_b$ to the expression $(F_b - F_h)/F_s$. The nodes highlighted in yellow are different than those in the first tree.}
        \label{fig:mutation}
    \end{subfigure}
    
    \begin{subfigure}[b]{\linewidth}
        \centering
        \includegraphics[width=0.7\linewidth]{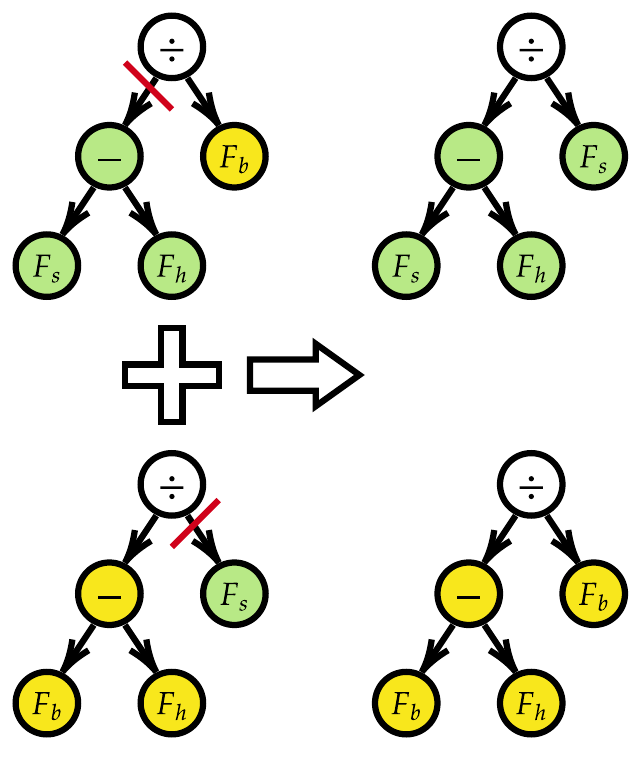}
        \caption{Crossover from the expressions $(F_s - F_h)/F_b$ and $(F_b - F_h)/F_s$. Two possible resulting expressions are $(F_s - F_h)/F_s$ and $(F_b - F_h)/F_b$.}
        \label{fig:crossover}
    \end{subfigure}

    \caption{Examples of mutation and crossover when applying \ac{GP} to \ac{SR}}
    \label{fig:genetic_programming}
\end{figure}

Finding non-linear mathematical expressions that approximate data requires an efficient search strategy. For this reason, we use \texttt{PySRRegressor}\footnote{\url{https://astroautomata.com/PySR/api/}} to perform a heuristic search based on \ac{GP} \citep{banzhaf_genetic_2000, mei_explainable_2023}. \ac{GP} relies on three fundamental operations on the trees:

\begin{itemize}
    \item Mutation: a tree is modified randomly (Fig. \ref{fig:mutation})
    \item Crossover (or recombination): two trees are combined to create two new trees (Fig. \ref{fig:crossover})
    \item Selection: the best trees are selected
\end{itemize}

During the search for optimal trees, their quality is assessed using \ac{MSE}. For the values $l_i$ in the latent dimension $L_i$ and the corresponding values $e_i$ generated by a tree's expression, \ac{MSE} is calculated as follows:

\begin{equation}
    \ac{MSE} =  \frac{1}{n}\sum_{i=1}^n (e_i - l_i)^2
\end{equation}

A \ac{GP} algorithm consists of four steps:

\begin{enumerate}
    \item Initialization: generate an initial population of $N_t$ random trees, where each tree represents a random physical expression.
    \item Selection: evaluate the candidate trees using the \ac{MSE} between the expression represented by the tree and the latent dimension data. Retain only the top $N_b$ trees with the lowest \ac{MSE} values.
    \item Evolution: apply genetic operations (mutation, crossover and selection) to the population of trees over multiple iterations to refine their quality.
    \item Termination: stop the evolutionary process once a termination condition is satisfied (here, a predefined number of iterations)
\end{enumerate}

As a result, the trees progressively improve with every iteration, producing easily interpretable physical formulas that more accurately approximate the latent space dimensions.

In this work, the maximum depth of the trees is $d = 5$ (to limit the number of variables to 3 and the number of operators to 2), the maximum complexity of the expressions is $c_{max} = 7$, and the number of iterations before termination is $N_{it} = 400$. The best model is the one with the lowest \ac{MSE}. All other hyperparameters are kept at their default values.

\section{Results}
\label{app:results}

\subsection{Clustering}
\label{app:results_clustering}

\begin{figure*}[htbp]
    \centering
    \begin{subfigure}[b]{\textwidth}
        \centering
        \includegraphics[width=\textwidth]{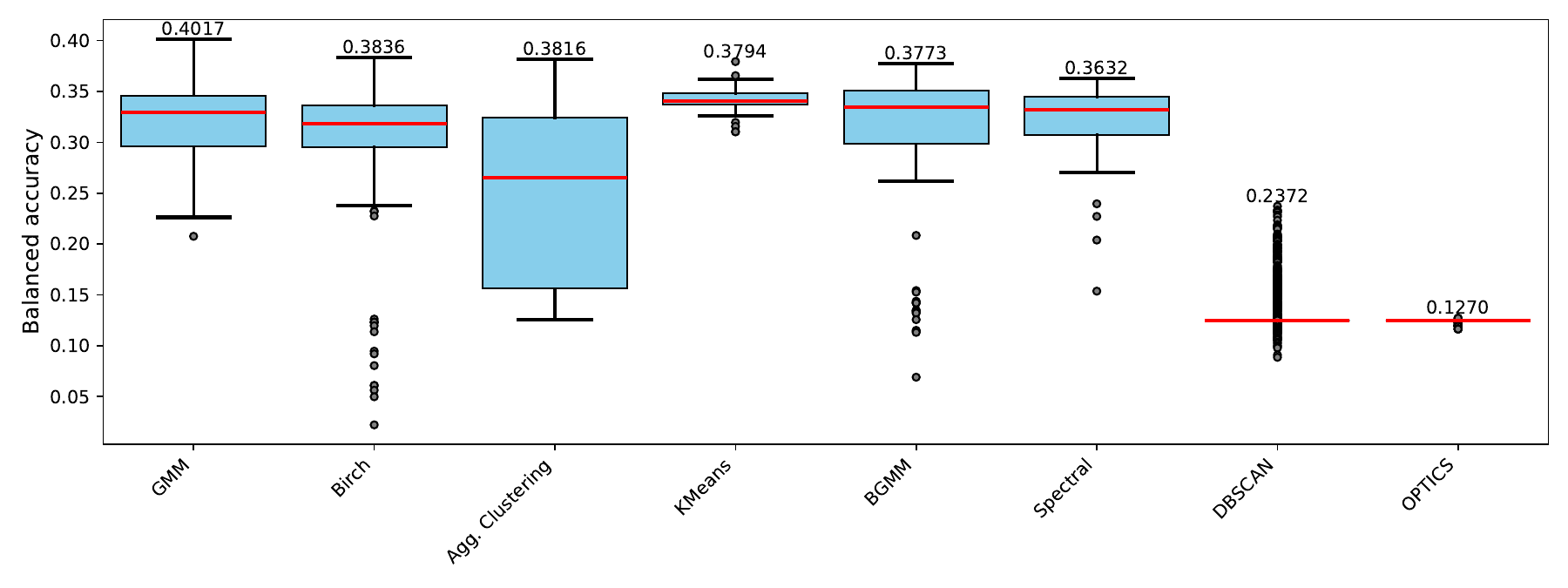}
        \caption{Clustering algorithm}
        \label{fig:clustering_alg}
    \end{subfigure}
    
    \begin{subfigure}[b]{0.24\textwidth}
        \centering
        \includegraphics[width=\textwidth]{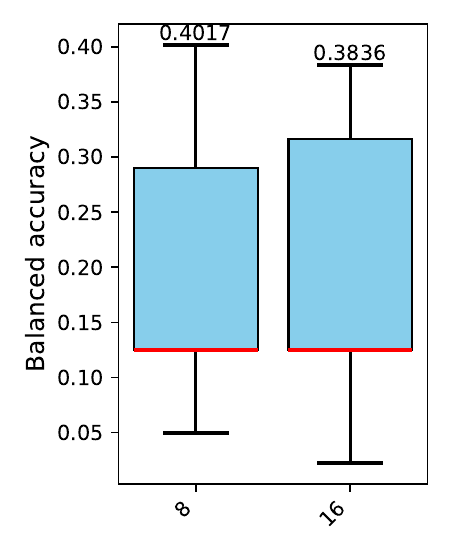}
        \caption{Embedding dimension}
        \label{fig:embed}
    \end{subfigure}
    \hfill
    \begin{subfigure}[b]{0.24\textwidth}
        \centering
        \includegraphics[width=\textwidth]{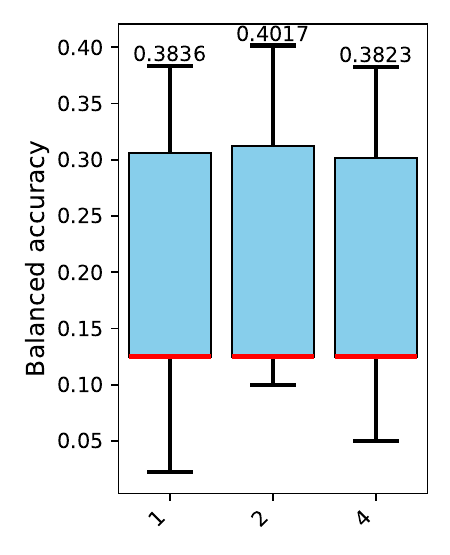}
        \caption{Number of layers}
        \label{fig:layers}
    \end{subfigure}
    \hfill
    \begin{subfigure}[b]{0.24\textwidth}
        \centering
        \includegraphics[width=\textwidth]{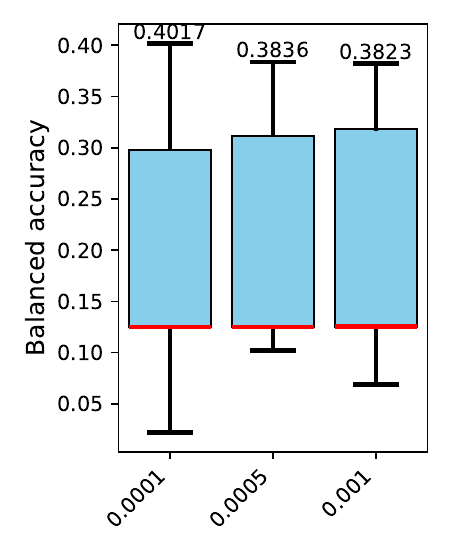}
        \caption{\ac{LR}}
        \label{fig:lr}
    \end{subfigure}
    \hfill
    \begin{subfigure}[b]{0.24\textwidth}
        \centering
        \includegraphics[width=\textwidth]{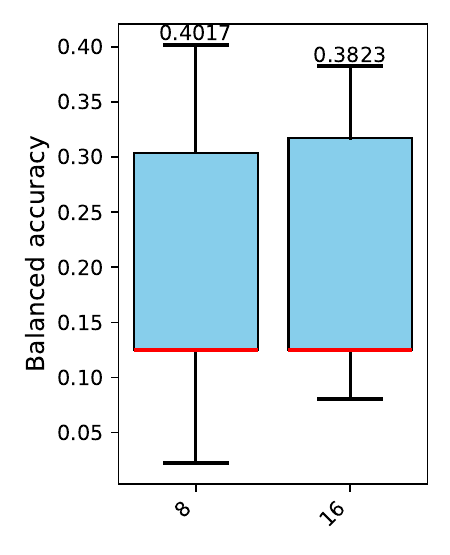}
        \caption{Number of heads}
        \label{fig:nhead}
    \end{subfigure}
    
    \caption{The impact of different autoencoder hyperparameters and clustering algorithms on the balanced accuracy. The red horizontal bars indicate the median values, and the black horizontal bars that delimit the light blue region indicate the first and the third quartiles. The clustering algorithm box plots are sorted by decreasing maximum balanced accuracy.}
    \label{fig:box_plots}
\end{figure*}

Figure \ref{fig:box_plots} summarizes the effect of the hyperparameters listed in Table \ref{tab:summary_autoencoder} and the clustering algorithms summarized in Table \ref{tab:summary_cluster} on balanced accuracy. 

All the figures use box plots to visualize the results. A box plot presents distributions using five main components: the median (the red line inside the box representing the 50th percentile), the lower quartile (Q1) and upper quartile (Q3) (i.e., the bottom and top lines of the box, corresponding to the 25th and 75th percentiles), the \ac{IQR} (i.e., the difference between Q3 and Q1), the whiskers (lines extending from the box to the smallest and largest values within 1.5 $\times$ \ac{IQR} from Q1 and Q3), and the outliers (individual points beyond the whiskers). In each figure, the box plots show the distribution of performance metrics when only one hyperparameter is fixed at a specific value, while all other hyperparameters are allowed to vary freely.

The clustering algorithm choice introduces the most significant variations in balanced accuracy. The remaining hyperparameters pertain to the autoencoder and have a smaller impact on balanced accuracy:

\begin{itemize}
\item Embedding dimensions: using 8 embedding dimensions results in a small latent space, which provides the best performance. This outcome indicates that the most relevant spectral features are captured in a smooth spectrum, while finer-grained structures are less significant.
\item Number of layers: the optimal architecture consists of 2 layers, balancing the simplicity of a single-layer model and the potential overfitting risks of a 4-layer model. 
\item \ac{LR}: a \ac{LR} of $10^{-4}$ yields the best balanced accuracy. Large rates lead to worse latent representations due to overfitting \citep{li_research_2019} and ineffectiveness in optimization \citep{peng_comprehensive_2024}.
\item Number of heads: using 8 heads is preferable as the autoencoder captures essential features in the latent space while avoiding focusing on irrelevant or noisy features.
\end{itemize}

Some clustering algorithms allow the number of clusters to be explicitly specified, unlike \ac{DBSCAN} and \ac{OPTICS}, which determine clusters based on density. Algorithms that allow one to define the number of clusters have better performance compared to those that rely on density-based clustering methods. Among all, \ac{GMM} achieves the highest balanced accuracy ($\approx$ 40\%). KMeans also performs well, achieving a balanced accuracy of 37.94\% (approximately 2\% lower than \ac{GMM}) while being the least sensitive to variations in other hyperparameters. Overall, the choice of the clustering algorithms' hyperparameters introduces a significant variation in balanced accuracy, compared with the uncertainty computed on the highest balanced accuracy ($40.5_{-4.2}^{+4.4} \%$ for a 99\% confidence interval around the median). Simpler configurations perform better because they avoid overfitting and effectively capture the most relevant spectral information.

\begin{figure}
    \centering
    
    \begin{subfigure}[b]{0.24\textwidth}
        \centering
        \includegraphics[width=\textwidth]{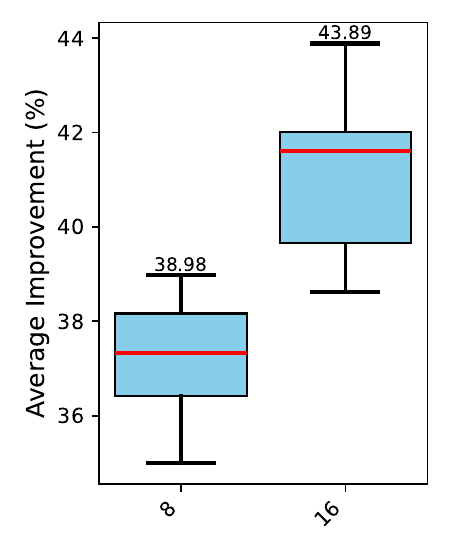}
        \caption{Embedding dimension}
        \label{fig:regression_embed}
    \end{subfigure}
    \hfill
    \begin{subfigure}[b]{0.24\textwidth}
        \centering
        \includegraphics[width=\textwidth]{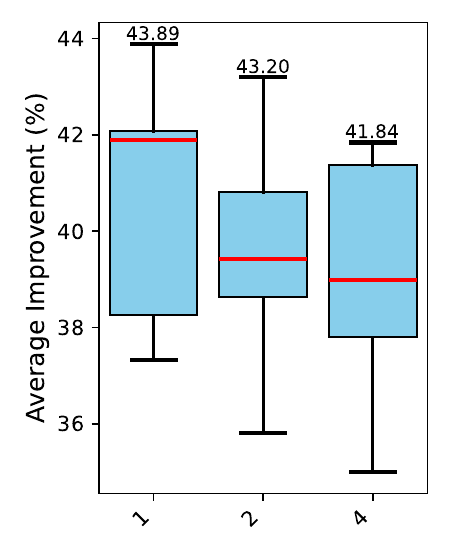}
        \caption{Number of layers}
        \label{fig:regression_layers}
    \end{subfigure}
    \hfill
    \begin{subfigure}[b]{0.24\textwidth}
        \centering
        \includegraphics[width=\textwidth]{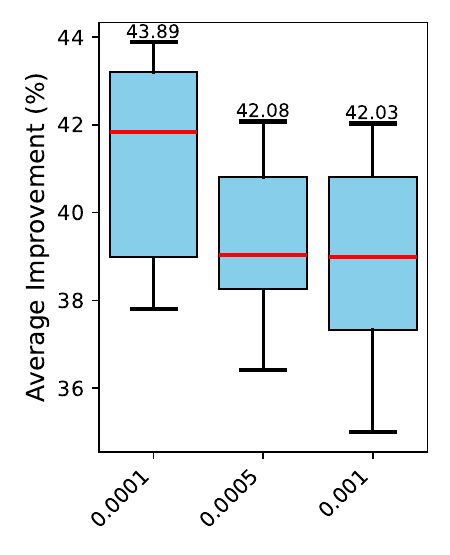}
        \caption{\ac{LR}}
        \label{fig:regression_lr}
    \end{subfigure}
    \hfill
    \begin{subfigure}[b]{0.24\textwidth}
        \centering
        \includegraphics[width=\textwidth]{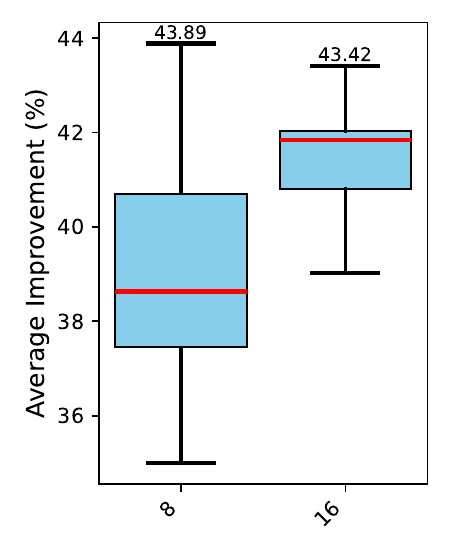}
        \caption{Number of heads}
        \label{fig:regression_nhead}
    \end{subfigure}
    
    \caption{The impact of different autoencoder hyperparameters and regression algorithm on the average \ac{MAE} improvement over the baseline. The red horizontal bars indicate the median values, and the black horizontal bars that delimit the light blue region indicate the first and the third quartiles. }
    \label{fig:box_plots_regression}
\end{figure}

\subsection{Regression}
\label{app:results_regression}

Figure~\ref{fig:box_plots_regression} summarizes the influence of the hyperparameters listed in Table~\ref{tab:summary_autoencoder} on the average \ac{MAE} improvement across all predicted physical quantities, with respect to a mean baseline (i.e., predicting the mean value of the training set). The embedding dimension shows the most significant impact, with an improvement of $\approx5\%$ when increasing the dimension from 8 to 16.

This result is in contrast with that of the clustering scenario, where smaller embedding dimensions lead to better results. In regression, a higher-dimensional latent space retains more details, which are relevant to the prediction of physical quantities. This representation supports fine-grained reconstructions that benefit the downstream regression task. Clustering, instead, benefits from smaller embeddings, which emphasize coarse-grained features and neglect irrelevant detail.

\end{appendix}
\end{document}